\documentclass[%
  reprint,
  a4paper,
  aps,
  prd,
  superscriptaddress,
  nofootinbib,
  amsmath,amssymb
]{revtex4-2}

\usepackage[a4paper,margin=2cm]{geometry}

\usepackage{amsmath}
\usepackage{amssymb}
\usepackage[bottom]{footmisc}
\usepackage{braket}
\usepackage{slashed}
\usepackage{verbatim}
\usepackage{adjustbox}
\usepackage{enumitem} 
\usepackage[dvipsnames]{xcolor}
\usepackage{mathtools}

\usepackage{xcolor}
\usepackage{bm,amsmath,amssymb}
\usepackage[utf8]{inputenc}
\usepackage{comment}
\usepackage{footmisc}
\usepackage{diagbox}
\usepackage{amsmath}
\usepackage{cancel}
\usepackage{physics}
\usepackage{dcolumn}
\usepackage{bm}

\usepackage{graphicx}
\usepackage{dcolumn}
\usepackage{bm}
\usepackage{physics}
\usepackage{amsmath}
\usepackage{bbold}
\usepackage{comment}
\usepackage[utf8]{inputenc}
\usepackage{xcolor}
\usepackage{amssymb}
\usepackage{graphicx}
\usepackage[T1]{fontenc}
\usepackage{makecell}

\usepackage[colorlinks=true, urlcolor=blue, linkcolor=blue, citecolor=red, hyperindex=true, linktocpage=true]{hyperref}

\def\CA{\mathcal{A}}
\def\CB{\mathcal{B}}

\def\CE{\mathcal{E}}
\def\CF{\mathcal{F}}

\def\CK{\mathcal{K}}

\def\CM{\mathcal{M}}

\def\CO{\mathcal{O}}

\begin{document}

\title{Duality and four-dimensional black holes:\\ gravitational waves, algebraically special solutions, pole skipping,\\ and the spectral duality relation in holographic thermal CFTs}

\author{Sa\v{s}o Grozdanov}
\affiliation{Higgs Centre for Theoretical Physics, University of Edinburgh, Edinburgh, EH8 9YL, Scotland, 
}
\affiliation{Faculty of Mathematics and Physics, University of Ljubljana, Jadranska ulica 19, SI-1000 Ljubljana, Slovenia
}

\author{Mile Vrbica}
\affiliation{Higgs Centre for Theoretical Physics, University of Edinburgh, Edinburgh, EH8 9YL, Scotland,
}


\begin{abstract}
The physics of gravitational waves and other classical fields on specifically four-dimensional backgrounds of black holes exhibits electric-magnetic-like dualities. In this paper, we discuss the structure of such dualities in terms of geometrical quantities with a physically-intuitive interpretation. In turn, we explain the interplay between the algebraic structure of black hole spacetimes and their associated dualities. For large classes of black hole geometries, explicit constructions are presented. We then use these results and apply them to the holographic study of three-dimensional conformal field theories (CFTs), discussing how such dualities place stringent constraints on the thermal spectra of correlators. In particular, the dualities enforce the recently-developed spectral duality relation along with a multitude of implications for the physics of thermal CFTs. A number of numerical results supporting our conclusions is also presented, including a demonstration of how the longitudinal spectrum of quasinormal modes determines the transverse spectrum, and vice versa.
\end{abstract}

\maketitle
\tableofcontents

\section{Introduction}

The physics of black holes in $3+1$ dimensions is interesting and important from a theoretical, observational and experimental point of view. One reason is the overwhelming evidence for their existence in our 4$d$ universe (see e.g. \cite{Ghez:2008ms,Gillessen:2008qv,EventHorizonTelescope:2019dse}). Another is the intricate connection that 4$d$ gravity has to quantum field theory in 3$d$ as a result of the gauge-gravity duality or holography (see e.g.~\cite{Zaanen:2015oix,Ammon:2015wua,Hartnoll:2018xxg}). Four-dimensional gravity is not only relevant and essential for understanding of the physics in our universe, it is also special. In pure gravity, it is the lowest dimension that supports the existence of propagating gravitational spin-$2$ degrees of freedom. Moreover, in 4$d$, gravitational fluctuations (gravitational waves) enjoy special symmetries absent in higher dimensions. Such symmetries are most easily understood as being analogous to the electric-magnetic duality of pure Maxwell theory in 4$d$. This is related to the fact that, in 4$d$, spin-$1$ and spin-$2$ gauge fields have exactly two polarisations each.

In this paper, we continue the exploration of constraints of such 4$d$ symmetries on the physics of gravitational and Maxwell fields, expanding on the results of Ref.~\cite{Grozdanov:2024wgo}, which focused on their holographic interpretation and implications for the thermal spectra of dual 3$d$ conformal field theories. As already mentioned, such dualities are reminiscent of the Maxwell electric-magnetic duality where the vacuum equations of motion and the Bianchi identities exhibit the same form---they are dual to each other. In the context explored here, such dualities were originally discovered by Chandrasekhar and Detweiler \cite{Chandrasekhar:1975zza} (see also Ref.~\cite{Chandrasekhar:1985kt}) to demonstrate the isospectrality of the quasinormal modes (QNMs) between the even and the odd channels of gravitational waves on Schwarzschild backgrounds. Notably, such dualities act only on the linearised perturbations of the fields rather than on the background fields themselves. As such, their structure is much more subtle, and, in standard formulations, parallels the formalism of supersymmetric quantum mechanics through the structure of Darboux transformations \cite{Anderson:1991kx,Glampedakis:2017rar,Lenzi:2021njy,Grozdanov:2023txs}. Such dualities are also related to integrable structures used, for example, to compute the black hole greybody factors \cite{Lenzi:2022wjv,Lenzi:2023inn,Jaramillo:2024qjz}.

In this paper, we focus on the geometric interpretation of such dualities and formulate them in a way which lends itself to a physical interpretation both in terms of the physics of gravitational waves and in their dual holographic settings. We focus on the neutral Schwarzschild black hole and black brane, and generalise our discussion to a charged (Reissner–Nordstr\" om) black hole as well as the linear axion model in asymptotically anti-de Sitter space \cite{Andrade:2013gsa}. We present, in detail, how dualities act in these cases, and comment on the extent of generality of the results. Furthermore, we discuss the `self-dual' limit, in which the dualities reduce to the standard electric-magnetic self-duality, be it for the Maxwell field (also known as the S-duality) or the linearised metric (see e.g.~Ref.~\cite{Hull:2001iu}).

An essential part of this story is the realisation that the duality maps between different channels of lineairsed perturbations are not always invertible. This fact is related to the existence of the algebraically special solutions of the perturbed Einstein's equations (see e.g.~\cite{Chandrasekhar:1985kt,Stephani:2003tm}). We elaborate on this interplay and offer a pedagogical introduction to the algebraic classification of spacetimes. We recap the main well-known results, and apply them to the theory of linearised perturbations. In that setting, the algebraic classification is a subtle problem that has, to our knowledge, been seldom addressed in literature (though discussions have appeared for example in Refs.~\cite{Couch:1973zc, Chandrasekhar:1984mgh, Dain:2000wuq, Dias:2013hn,Araneda:2015gsa}). We comment on these subtleties at length and present a simple method for computing the so-called \emph{algebraically special frequencies}, which parametrise and fix the duality equations. We present closed-form expressions for all discussed examples, which we also believe should be useful to future explorations. On the way, we discuss the relation between the algebraically special frequencies and pole skipping \cite{Grozdanov:2017ajz,Blake:2017ris,Blake:2018leo} \cite{Grozdanov:2019uhi,Blake:2019otz}. The pole-skipping discussion is both an elaboration and a generalisation of our results from Ref.~\cite{Grozdanov:2023txs}.

In asymptotically anti-de Sitter space, the dualities do not yield the isospectrality of the QNMs. Instead, a more subtle structure is present, manifest in the \emph{spectral duality relation}, introduced in our recent work \cite{Grozdanov:2024wgo}. Here, we expand on and utilise the structures imposed by this relation in 4$d$ theories of gravity, electromagnetism and other fields in spaces with black holes and a negative cosmological constant, as well as their dual 3$d$ conformal field theories (CFTs). From the holographic point of view, we study the thermal (quasinormal mode) spectra \cite{Kovtun:2005ev} of both the energy-momentum and conserved current correlators. The best understood example of such a CFT is the worldvolume theory of a large stack of M2 branes, or the ABJM theory \cite{Aharony:2008ug}. With our construction, we then focus on the manifestation of the bulk dualities on boundary retarded correlators and comment on their universal form in CFTs with a large number of colours or a large central charge---the so-called large-$N$ theories. Using the thermal product formula recently developed in Ref.~\cite{Dodelson:2023vrw}, we rederive the spectral duality relation \cite{Grozdanov:2024wgo}, and show how it precisely relates the longitudinal (even) and transverse (odd) spectra. In particular, we present explicit constructions that can be used to compute the entire spectrum of the transverse channel only from the knowledge of the longitudinal spectrum. In an analogous manner, the longitudinal spectrum follows from the transverse spectrum and a single wavevector-dependent function, which must be determined by other means. On the way, we rearticulate several well-known older results \cite{Herzog:2007ij,Bakas:2009pbm,Davison:2014lua,deHaro:2008gp}, which can all be understood from the `self-dual' limits of our results. In all of the {\em non-self-dual} cases, the central role is played by the algebraically special frequencies, which are the main input from the bulk physics. We also discuss the role of the algebraically special solutions on the boundary and remark on the `universal' way in which they control the dualities. Moreover, we present several numerical results, including a simple approximation for the (all-order) dispersion relation of the diffusive mode through the knowledge of the hydrodynamic sound mode, the algebraically special frequency and the asymptotic large-frequency WKB data. The discussion of thermal spectra in this paper is complementary to the one present in Ref.~\cite{GV:2025}, where we discuss the spectral duality relation in a more general setting of large-$N$ field theories in any number of dimensions.

This paper is structured as follows: in Section~\ref{sec:setup}, we introduce the general 4$d$ geometrical setup and introduce the relevant notation. We discuss the linearisation of the metric and the Maxwell field around the set of maximally symmetric black hole backgrounds with no reference to a specific action. We introduce convenient electric and magnetic variables and offer their physical interpretation. In section~\ref{sec:algebraical}, we then present a pedagogical summary of the important aspects of the algebraic classification of spacetimes and discuss various subtleties that arise in linearised theories. We also present an ansatz that is particularly convenient for the analysis of the algebraically special solutions on black hole backgrounds. In Section~\ref{sec:dualities}, we consider the Einstein-Maxwell action and present the general structure of dualities with concrete examples. We demonstrate how, in the case of a Schwarzschild black hole, the `meaning' of dualities can be understood in a rather transparent way when expressed in terms of variables that have a direct physical interpretation.  Finally, in Section~\ref{sec:AdSCFT}, we discuss a number of applications of our results to the holographic analysis of the spectra of thermal two-point correlation functions.

\section{The setup}
\label{sec:setup}
\subsection{Curvature tensors and their decomposition}
\label{sec:CurvatureTensorsAndDecomposition}
In this section, we introduce the necessary machinery that will allow us to treat both the physics of gravitational waves and holography in a way that is naturally suited to the dualities considered in this work. Although we will ultimately be interested in linearised perturbations around maximally symmetric black hole backgrounds, we leave the construction of this section general by not restricting ourself to linearised gravity or specific geometries. This is to encourage possible future generalisations of the contents of this paper. Note that, however, most of our present discussion is only applicable to 4$d$ (bulk) spacetimes.

The main object of interest will be the Weyl tensor. This is because, in the context of dualities discussed here, it can thought of as the gravitational analogue of the Maxwell field strength tensor. The discussion will rely heavily on the spacetime having two preferred null directions, or equivalently, a preferred timelike direction (such as the timelike Killing vector of a stationary spacetime), and a preferred spacelike direction (such as the radial direction). We work with a mostly plus metric signature and with the Riemann tensor defined as $[\nabla_a, \nabla_b]V^c=R^{c}{}_{dab}V^d$. We mostly follow the discussion of Ref.~\cite{Stephani:2003tm}.

We begin by introducing a pair of orthogonal vectors: a normalised spacelike vector $n^a$ and a normalised timelike vector $u^a$, so that
\begin{subequations}
\label{def:u&n}
\begin{align}
    n_a n^a&=1,\\
    u_a u^a&=-1,\\
    n_a u^a&=0.
\end{align}
\end{subequations}
This allows us to introduce two null vectors $\ell^a_\pm$,
\begin{equation}
    \ell^a_\pm=\frac{1}{\sqrt{2}}(u^a \pm n^a), \label{def:null_vectors}
\end{equation}
normalised to $\ell^a_+ \ell^b_- g_{ab}=-1$. The induced metric on the $n$-orthogonal hypersurface is then 
\begin{equation}
    h_{ab}=g_{ab}-n_a n_b, \label{def:induced_metric}
\end{equation}
and the projector to the space orthogonal to both $u_a$ and $n_a$ is
\begin{equation}
    \Delta_{ab}=h_{ab}+u_a u_b=g_{ab}+2 \ell^+_{(a}\ell^-_{b)}. \label{def:projectors}
\end{equation}
$\Delta_{ab}$ is traceless with respect to the metric and the induced metric \eqref{def:induced_metric}, and is orthogonal to $u_a$, $n_a$, and also $\ell_\pm^a$. Using this projector, we can define the projection of a vector onto the submanifold orthogonal to $u_a$ and $n_a$ as
\begin{equation}
    A_{\expval{a}}=\Delta_{a}{}^b A_b, \label{def:transverse_vector}
\end{equation}
and the traceless projection of a tensor as
\begin{equation}
    A_{\expval{ab}}=\Delta_a {}^c \Delta_b {}^d A_{(cd)}-\frac{1}{2}\Delta_{ab}\Delta^{cd}A_{cd}, \label{def:transverse_tensor}
\end{equation}
with $A_{\expval{ab}}$ traceless and orthogonal to both $u_a$ and $n_a$. As such, it can effectively be represented by a $2\times 2$ matrix with two independent elements.

Consider first the familiar case of the electromagnetic field strength tensor $F_{ab}$ defined in terms of the Maxwell field $A_a$ as
\begin{equation}
    F_{ab}=2 \partial_{[a}A_{b]},
\end{equation}
with its Hodge dual
\begin{equation}
    \widetilde{F}_{ab}=\frac{1}{2}\epsilon_{abcd}F^{cd},
\end{equation}
where $\epsilon_{abcd}$ is the Levi-Civita tensor. We decompose $F_{ab}$ into its electric and magnetic parts with respect to the spacelike vector $n^a$ by defining
\begin{subequations}
\begin{align}
    E_a&= F_{ab} n^b,\label{def:Emaxwell}\\ 
    B_a&= \widetilde{F}_{ab}n^b, \label{def:Bmaxwell}
\end{align}
\end{subequations}
where both $E_a$ and $B_a$ have $3$ independent components each, accounting for the $6$ independent components of the field strength tensor, which can be expressed as
\begin{equation}
    F_{ab}=2 E_{[a}n_{b]}-\epsilon_{abcd}B^c n^d.
\end{equation}
The dual tensor $\widetilde{F}_{ab}$ is acquired by substituting $F\rightarrow \widetilde{F}$, $E\rightarrow B$ and $B \rightarrow -E$ in the equation above.

An analogous situation in gravity can be constructed in terms of the curvature tensors, in particular, the Weyl tensor $C_{abcd}$ defined as the traceless part of the Riemann tensor
\begin{equation}
    C^{ab}{}_{cd}= R^{ab}{}_{cd}-4 S^{[a}{}_{[c}\delta^{b]}{}_{d]},
\end{equation}
where $S_{ab}$ is the Schouten tensor
\begin{equation}
    S_{ab} = \frac{1}{2}\qty(R_{ab}-\frac{1}{6}R g_{ab}).
\end{equation}
The Weyl tensor contains the information about the Riemann tensor that is not directly determined by the Ricci tensor. This makes it a relevant quantity in the study of gravitational waves. We can also introduce the dual Weyl tensor
\begin{equation}
    \widetilde{C}^{ab}{}_{cd}= \frac{1}{2}\epsilon^{abef}C_{efcd}, \label{def:weyl_dual}
\end{equation}
where we note that $\epsilon^{abef}C_{efcd}=C^{abef}\epsilon_{efcd}$. The Weyl tensor can be decomposed into its electric and magnetic components by defining
\begin{subequations}
\label{def:EBWeyl}
\begin{align}
    E_{ab}&=C_{acbd}n^c n^d,\label{def:Eweyl}\\
    B_{ab}&=\widetilde{C}_{acbd}n^c n^d,\label{def:Bweyl}
\end{align}
\end{subequations}
where $E_{ab}$ and $B_{ab}$ are both symmetric, traceless, and have $5$ independent components each, together accounting for the $10$ independent components of the Weyl tensor, which can now be written as
\begin{align}
    -\frac{1}{2}C^{ab}{}_{cd}=&\,4 n^{[a}E^{b]}{}_{[c}n_{d]}+2\delta^{[a}{}_{[c}E^{b]}{}_{d]}+\\&+ \epsilon^{abef}n_e n_{[c}B_{d]f}+\epsilon_{cdef}n^e n^{[a}B^{b]f}. \nonumber
\end{align}
The expression for the dual Weyl tensor is acquired by substituting $C \rightarrow \widetilde{C}$, $E\rightarrow B$ and $B \rightarrow -E$, which is analogous to the situation in electromagnetism.

We have chosen to define the electric and magnetic tensors with respect to an $n$-orthogonal foliation, which is to be contrasted with the more usual case of choosing a timelike vector for that role. This, it turns out, is more convenient for the holographic analysis in Section~\ref{sec:AdSCFT}. For all other purposes, $n^a$ can be replaced with $u^a$, with only differences appearing in possible minus signs.

It is important to note that the decomposition of the curvature tensor into its electric and magnetic components has a direct interpretation in terms of gravitational waves (see Section~\ref{subsec:gravitational_waves}), holography (see Section~\ref{sec:AdSCFT}), as well as the non-linear notions of tidal forces and frame-dragging \cite{Nichols:2011pu}.

\subsection{The background}
In this paper, we will focus on the geometries of maximally symmetric (i.e., non-rotating) black holes in four spacetime dimensions. More concretely, we consider a product manifold $\CM \times \CK$, where $\CM$ is $2$-dimensional Lorentzian space described by coordinates $(t,r)$, and $\CK$ is a $2$-dimensional Riemannian maximally symmetric space described by coordinates $(\chi,\phi)$. This means that the slices of constant $t$ and $r$ are maximally symmetric spaces. A class of such spacetimes is described by the metric
\begin{equation}
    ds^2=-f(r)dt^2+\frac{dr^2}{f(r)}+r^2 \gamma_{AB}dx^A dx^B , \label{def:background_metric}
\end{equation}
where
\begin{equation}
    \gamma_{AB}dx^A dx^B = \frac{d\chi^2}{1-K \chi^2}+\chi^2 d\phi^2,
\end{equation}
with the capital Latin letters representing the coordinates of $\CK$. Here, $K$ is the normalised sectional curvature that parametrises the geometry of the horizon(s): $K=1$ yields the standard spherical horizon, $K=0$ yields a planar horizon (i.e., a black brane) and $K=-1$ a hyperbolic horizon. The space is equipped with the Levi-Civita tensor, which we define to obey
\begin{equation}
    \epsilon_{tr\chi\phi}=\sqrt{-g}.
\end{equation}

The solutions to $f(r)=0$ indicate a horizon. We limit our discussion to the region of spacetime outside the outer event horizon (which we denote by $r_0$), and, possibly, within the cosmological horizon, where $f(r)>0$. In this region, we introduce the tortoise coordinate $r_*$ through
\begin{equation}
    \frac{dr_*}{dr}=\frac{1}{f(r)}. \label{def:tortoise}
\end{equation}
The Hawking temperature associated with the event horizon is
\begin{equation}
    T=\frac{f'(r_0)}{4\pi} \label{def:hawking}.
\end{equation}
We also define the timelike Killing and the spacelike radial vectors from Eq.~\eqref{def:u&n} as
\begin{subequations}
\begin{align}
    u^a \partial_a &=\frac{1}{\sqrt{f(r)}} \partial_t, \\
    n^a \partial_a &=\sqrt{f(r)} \partial_r.
\end{align}
\end{subequations}
This gives rise to the \emph{outgoing} $\ell_+^a$ and \emph{ingoing} $\ell_-^a$ null vectors through the definition \eqref{def:null_vectors}, pointing away and towards the black hole, respectively. The projector $\Delta_{ab}$, as defined in Eq.~\eqref{def:projectors}, then projects onto the slices of constant $t$ and $r$.

The class of metrics \eqref{def:background_metric} has a non-zero electric part of the Weyl tensor $E_{ab}$, while the magnetic part $B_{ab}$ vanishes. Specifically, the trace $\Delta^{ab}E_{ab}$, which represents the `Coloumb-like' component of the gravitational field, is finite, while the off-diagonal parts vanish (see Appendix~\ref{app:NP}).

One can define the (scalar) eigenfunctions of the Laplace-Beltrami operator on $\CK$ as
\begin{equation}
    \left(D_A D^A +\mu\right)Y^{(\mu)}(\chi,\phi)=0, \label{def:harmonics}
\end{equation}
where $D_A$ is the covariant derivative compatible with the metric $\gamma_{AB}$. In the $K=1$ case, $Y$ are the standard spherical harmonics, with $\mu=l(l+1)$, where $l$ is the representation index of SO$(3)$. In the $K=0$ case, $\mu=k^2$, with $k$ the Euclidean wavevector, and $Y$ are plane waves expressed in polar coordinates. The $K=-1$ is similar to the $K=1$ case, and the corresponding eigenfunctions $Y$ are hyperbolic harmonics (see e.g.~Ref.~\cite{helgason2000groups}). We will henceforth suppress the superscript $\mu$. The eigenfunction $Y$ is parity-even and allows us to introduce a parity-even vector and a parity-even symmetric traceless tensor on $\CK$:
\begin{subequations}
\begin{align}
    Y_A&=D_A Y, \\
    Y_{AB}&=D_A D_B Y+\frac{1}{2}\mu \gamma_{AB} Y.
\end{align}
\end{subequations}
Analogously, we can introduce a parity-odd vector and a parity-odd symmetric traceless tensor $\CK$:
\begin{subequations}
\begin{align}
    X_A&=\epsilon_{A}{^B}Y_B,\\
    X_{AB}&=D_{(A}X_{B)}.
\end{align}
\end{subequations}
We have the identity $X_{AB}=\epsilon_A{}^C Y_{CB}$ relating the traceless tensors of both parities. Here, $\epsilon_{AB}$ is the Levi-Civita tensor on $\CK$. We will also use the trivial extension of those vectors and tensors on the entire 4$d$ space, e.g.,  $Y_{A}dx^A=Y_{a}dx^a$.

\subsection{The perturbations}
\label{subsec:perturbations}
We now linearly perturb the metric \eqref{def:background_metric} as $g_{ab}\rightarrow g_{ab}+\delta g_{ab}$, and decompose it with respect to the complete set of scalar, vector and tensor harmonics (see Appendix~\ref{app:master}). We also consider similar linearised perturbations of the Maxwell field $A_a \rightarrow A_a + \delta A_a$ and any other present fields, appropriately decomposed in a similar sense. We further exploit the static nature of the geometry \eqref{def:background_metric} and decompose all fields into temporal Fourier modes $e^{-i \omega t}$.

The linearised theory enjoys the linearised diffeomorphism invariance of the metric $g_{ab}\rightarrow g_{ab}+\nabla_a \xi_b+\nabla_b \xi_a$, as well as the gauge invariance of the Maxwell field $A_a \rightarrow A_a + \partial_a \alpha$. Note that while the field strength tensor $F_{ab}$ is invariant under the electromagnetic gauge transformations, the Weyl tensor $C_{abcd}$ is \emph{not} invariant under linearised diffeomorphisms in curved spacetimes.  We also perturb the normalised timelike and spacelike vectors $u^a \rightarrow u^a+\delta u^a$, $n^a \rightarrow n^a + \delta n^a$, so that the normalisation and orthogonality conditions \eqref{def:u&n} still hold. The `natural' quantities encoding the Maxwell fields $E_{\expval{a}}$ and $B_{\expval{b}}$ can then expand in the basis of $Y_a$ and $X_a$ as
\begin{subequations}
\label{def:CECB}
\begin{align}
    E_{\expval{a}}&=\qty[\CE_+(r) Y_a+\CE_-(r) X_a]e^{-i\omega t}, \\
    B_{\expval{a}}&=\qty[\CB_-(r) Y_a+\CB_+(r) X_a]e^{-i\omega t}. 
\end{align}    
\end{subequations}
On the other hand, an analogous description of the metric perturbations is expressed in terms of $E_{\expval{ab}}$ and $B_{\expval{ab}}$, which we expand in the basis of the traceless tensors $Y_{ab}$ and $X_{ab}$ as
\begin{subequations}
\label{def:EB}
\begin{align}
    E_{\expval{ab}}&=\qty[E_+(r) Y_{ab}+E_-(r) X_{ab}]e^{-i\omega t}, \\
    B_{\expval{ab}}&=\qty[B_-(r) Y_{ab}+B_+(r) X_{ab}]e^{-i\omega t}.
\end{align}
\end{subequations}
The quantities denoted with $+$ are the even channel quantities (also called longitudinal, scalar or polar), while the quantities denoted with $-$ are the odd channel quantities (also called transverse, vector or axial). The quantities $E_\pm$ and $B_\pm$ are directly related to the complex Weyl scalars (see Appendix~\ref{app:NP}) and are relevant for several reasons. Firstly, they are linearly small by construction in that their background values are zero. Secondly, they are gauge invariant---they are independent of the perturbed values of $u^a$ and $n^a$ as well as the U($1$) gauge transformations and the linearised diffeomorphisms. Thirdly, they are the natural variables for diagnosing the algebraically special solutions (see Section~\ref{sec:algebraical}) and the natural variables on which the duality maps in Section~\ref{sec:dualities} act. Furthermore, they have a direct physical interpretation in terms of gravitational waves in regions of space where plane waves $e^{\pm i\omega r_*}$ are a good approximation to the solution of the Einstein's equations (see Section~\ref{subsec:gravitational_waves}). Finally, they have a direct interpretation in terms of the thermal CFT observables in AdS/CFT (see Section~\ref{sec:AdSCFT}). We conclude this section by noting that we could obtain the same variables by defining $E_{ab}$ and $B_{ab}$ with respect to the alternative foliation using a timelike vector that was mentioned at the end of Section~\ref{sec:CurvatureTensorsAndDecomposition}. The reason is that $E_\pm$ and $B_\pm$ only depend on the choice of the two null vectors of Eq.~\eqref{def:null_vectors}.

\subsection{Interpretation of the variables in terms of gravitational waves}
\label{subsec:gravitational_waves}

Consider a region of the radial coordinate for which the solutions to the Einstein's equations are well approximated with the radial dependence $e^{\pm i \omega r_*}$. This can be the near-horizon region of a black hole or de Sitter space, or the asymptotically flat region of some asymptotically flat space. If the gravitational wave is ingoing, i.e., propagating in the direction of decreasing $r$, then the scalar quantities behave approximately as $e^{-i\omega(t+r_*)}$ and are approximately annihilated by $\ell^a_- \partial_a$. In the regions of the spacetimes where those approximations hold, we have
\begin{subequations}
\label{eq:ingoingoutgoing}
\begin{align}
    &r_*\rightarrow \pm \infty: &E_++B_+ &\approx 0, \\
    & & E_--B_- &\approx 0.
\end{align}
Conversely, if the wave is outgoing, i.e., if various scalars behave as $\sim e^{-i t (\omega-r_*)}$ and are approximately annihilated by $\ell^a_+ \partial_a$, then
\label{eq:outgoing}
\begin{align}
    &r_*\rightarrow \pm \infty: &E_+-B_+ &\approx 0, \\
    & & E_-+B_- &\approx 0.
\end{align}
\end{subequations}
In the relevant regions of the spacetimes where such approximations hold, we therefore identify $E_+-B_+$ as an ingoing even channel wave and $E_-+B_-$ as an ingoing odd channel wave. The same can be said about ingoing and outgoing electromagnetic waves, where we need to substitute $E_\pm \rightarrow \CE_\pm$ and $B_\pm \rightarrow \CB_\pm$ in Eqs.~\eqref{eq:ingoingoutgoing}.

\section{Algebraically special spacetimes}
\label{sec:algebraical}

The notion of an algebraically special spacetime presents a measure of the simplicity of a given spacetime that is a priori independent of its isometries. Instead, it relies on the null structure of the Weyl tensor. Historically, it has been an indispensable tool for finding exact solutions to the Einstein equations, for example, in Kerr's derivation of the rotating black hole solution \cite{Kerr:1963ud}. Furthermore, it offers insights into the behaviour of null geodesics via the Goldberg-Sachs theorem (see below), as well as the geometry of the null infinity through the peeling theorem \cite{Wald:1984rg}. In the Newman-Penrose formalism \cite{Newman:1961qr} (see Appendix~\ref{app:NP}), the algebraic structure of a spacetime can be used to greatly simplify the Einstein's equations and offer various theoretical insights, including on the topic of duality structures discussed in this paper.

The vast majority of known exact solutions to the Einstein equations in four spacetime dimensions, if not all, are algebraically special \cite{Stephani:2003tm}. This includes all spherically symmetric geometries, Kerr and Taub-NUT black holes, pp-waves, the FLRW space, as well as the large classes of Kerr-Schild and Robinson-Trautman metrics. Note that the notion of an algebraically special spacetime relates to the geometry itself with no reference to the action from which it is derived.

In this section, we discuss the algebraic classification (also known as the Petrov classification) of spacetimes, which turns out to be crucial for the discussion of dualities. While there exists a rich body of literature on the algebraic properties of spacetimes (we largely rely on Ref.~\cite{Stephani:2003tm} and references therein), we will assume that the reader may not be overly not familiar with these concepts, and will therefore attempt to introduce the subject in a pedagogical manner. In this section, we will keep the notion of a spacetime geometry very general (not necessarily having the form Eq.~\eqref{def:background_metric}), which will help us to make certain general statements that will prove helpful for the discussion of the subtleties that pertain to the algebraic properties of spacetimes in linearised gravity.

\subsection{Algebraic classification of spacetimes}
In four spacetime dimensions, there exist several equivalent ways of algebraically classifying a spacetime. Here, we follow Penrose \cite{Penrose:1960eq} and consider a non-zero null vector field $\ell_a$ so that $\ell_a \ell^a=0$. Such a vector is called a \emph{principal null direction} (PND) of the Weyl tensor if and only if
\begin{equation}
    \ell_{[e}C_{a]bc[d}\ell_{f]}\ell^b \ell^c=0. \label{def:PND}
\end{equation}
In 4$d$, there can be at most four such independent vectors at any given point. The PNDs do not need to be independent but can appear with various multiplicities. If a PND $\ell_a$ has the multiplicity of more than one, then it is called degenerate, and the following holds:
\begin{equation}
    C_{abc[d}\ell_{f]}\ell^b\ell^c=0. \label{def:degeneratePND}
\end{equation}
If a spacetime has at least one degenerate PND at every point, then it is called \emph{algebraically special}. The ways in which the PNDs are organised give rise to the Petrov classification. There are six Petrov types and if a spacetime is of a certain Petrov type at every point, then the spacetime itself can be assigned a Petrov type. In this paper, we are mostly interested in type-II (one doubly degenerate PND) and type-D (two doubly degenerate PNDs) spacetimes, the latter being a subset of the former. Specifically, the geometries considered in this paper \eqref{def:background_metric} have two doubly degenerate PNDs and are therefore of type-D. The two null directions are simply the ingoing $\ell^a_-$ and the outgoing $\ell^a_+$ geodesic tangents defined in Eq.~\eqref{def:null_vectors}.

There are several ways to determine whether a spacetime is algebraically special, which are equivalent to Eq.~\eqref{def:degeneratePND}. One condition that is particularly convenient for the purposes of this paper is the question of whether (see Appendix~\ref{app:NP})
\begin{equation}
    E_{\expval{ab}}=\pm \ell_+^i \ell_-^j \epsilon_{ija}{}^d B_{\expval{db}}, \label{eq:EBalgSpec}
\end{equation}
where $+$ indicates that $\ell_+^a$ is the degenerate PND, and $-$ indicates that $\ell_-^a$ is the degenerate PND. Computationally, the simplest tool for such analyses is often provided by the Newman-Penrose formalism (see Appendix~\ref{app:NP}), which also offers a simple recipe for finding all the PNDs explicitly.

A partial physical interpretation of a degenerate PND is given by the Goldberg-Sachs theorem, which states that for a class of Ricci tensors (including the Ricci tensor of the metric defined in \eqref{def:background_metric} and all Ricci-flat cases), the spacetime will necessarily be algebraically special if it admits a shear-free geodesic null congruence, that is, if the spacetime possesses a null vector $\ell^a_-$ (or equivalently, $\ell^a_+$), that is both everywhere tangent to a geodesic,
\begin{equation}
    \ell^c_- \ell^{[a}_-\nabla_c \ell^{b]}_- =0,
\end{equation}
and shear-free,
\begin{equation}
    \nabla^{\langle a}_{\vphantom{-}}\ell_-^{b\rangle}=0. \label{eq:shear-free}
\end{equation}
In this case, $\ell^a_-$ is a degenerate PND. In passing, we mention the existence of the Mariot-Robinson theorem, which says that a geodesic shear-free null congruence condition is satisfied only if the spacetime admits a `test' electromagnetic field which is both null ($F_{ab}\ell^a=\widetilde{F}_{ab} \ell^a=0$) and obeys the Maxwell equations ($\nabla^a F_{ab}=\nabla^a \widetilde{F}_{ab}=0$). This is again a statement about the geometry without a reference to a specific action.

\subsection{Algebraically special linearised spacetimes}
\label{subsec:algebraically_linear}
Next, we consider what happens to the PNDs when the metric is dynamically perturbed. Naively, one may expect that one can take the original degenerate null directions $\ell^a_+$ and $\ell^a_-$ and perturb them with some linear corrections $\delta \ell^a_+$ and $\delta \ell^a_-$. It turns out that, generically, this is not a well-posed problem and that the perturbed defining equation for PNDs \eqref{def:PND} does not necessarily admit a solution. This is illustrated most clearly in the case of linearised perturbations around some conformally flat space. Such a space has a vanishing Weyl tensor and therefore has no PNDs that could be perturbed. Nevertheless, rich physics of algebraically special solutions in linearised gravity around black hole backgrounds exists \cite{Couch:1973zc, Chandrasekhar:1984mgh,Dias:2013hn,Araneda:2015gsa}, with null vectors that satisfy Eq.~\eqref{def:degeneratePND}. In this paper, we take the following approach: Suppose we have a type-D background spacetime (i.e., a spacetime with two doubly degenerate PNDs). We linearly perturb the metric $g_{ab}\rightarrow g_{ab}+\delta g_{ab}$ and a PND $\ell^a \rightarrow \ell^a + \delta \ell^a$. Then, one of two things can happen:
\begin{enumerate}
    \item either Eq.~\eqref{def:PND} has no solutions, irrespective of the choice of $\delta \ell_a$,
    \item or $\ell_a$ solves Eq.~\eqref{def:PND}, irrespective of the choice of $\delta \ell_a$. In this case, one can always find $\delta \ell^a$ so that Eqs.~\eqref{def:degeneratePND} and $\eqref{eq:EBalgSpec}$ are satisfied as well. This means that $\ell^a$ is a degenerate PND, and the perturbed spacetime is algebraically special (type-II). \label{enum:criterion}
\end{enumerate}
This statement is proven in Appendix~\ref{app:NP}. Note that the Goldberg-Sachs theorem does not translate to the linearised theory \cite{Dain:2000wuq}.

We now focus on the example at hand and consider the background metric from Eq.~\eqref{def:background_metric}, which will be the main focus of study in this work. If either of $\ell_\pm^a$ is a PND of the perturbed metric, then it must be doubly degenerate and the spacetime is algebraically special by the criterion \ref{enum:criterion} above. We can connect this with a statement made in terms of the variables introduced in Section \ref{subsec:perturbations}. In particular, the perturbed metric will be algebraically special, in the sense describe above, if and only if one of the following is fulfilled:
\begin{itemize}
    \begin{subequations}
    \label{eq:perturbed_conditions}
    \item either
    \begin{equation}
    \label{eq:perturbed_ingoing}
        E_++B_+=E_--B_-=0,
    \end{equation}
    in which case, $\ell^a_-$ is a doubly degenerate PND (such solutions are ingoing),
    \item or
    \begin{equation}
        E_+-B_+=E_-+B_-=0,
    \end{equation}
    in which case, $\ell^a_+$ is a doubly degenerate PND (such solutions are outgoing).          
    \end{subequations}
\end{itemize}
This follows directly from the condition \eqref{eq:EBalgSpec} and the definitions \eqref{def:EB}. Note the similarity between these equations and equations \eqref{eq:ingoingoutgoing}. This is not coincidental, as algebraically special spacetimes can be interpreted as purely ingoing or outgoing waves in the regions outlined in Section~\ref{subsec:gravitational_waves}. The criterion \eqref{eq:perturbed_conditions}, however, does not depend on any wavelike interpretation of the perturbed geometry. This is especially relevant on the boundary of anti-de Sitter (AdS) space, where gravitational waves have no ingoing or outgoing interpretation. Given some equations of motion, the criterion \eqref{eq:perturbed_conditions} will only be fulfilled for some specific ($\mu$-dependent) choices of $\omega$, called the \emph{algebraically special frequencies}.

One can avoid solving the equations of motions with the constraints of Eqs.~\eqref{eq:perturbed_conditions} by using an appropriate ansatz. A large class of manifestly algebraically special type-II metrics (including all the common black hole metrics) was written by Timofeev \cite{Timofeev1996} (see also Refs.~\cite{Stephani:2003tm,ioannisTalk}). All such (ingoing) metrics that are linearly close to the background of Eq.~\eqref{def:background_metric} are expressed by (up to linearised diffeomorphisms)
\begin{align}
    \frac{\delta g_{tt}}{f(r)}&=f(r) \delta g_{rr}=\delta g_{tr}=\xi(r) Y e^{-i\omega(t+r_*)}, \nonumber\\ 
    \delta g_{AB}&=r^2 \zeta_+   \gamma_{AB}Ye^{-i\omega(t+r_*)},\nonumber\\
    \delta g_{rA}&=\left(\varphi_+(r) Y_A+\varphi_-(r) X_A\right)e^{-i\omega(t+r_*)},\nonumber\\
    \frac{\delta g_{tA}}{f(r)}&= g_{rA}+\zeta_- X_Ae^{-i\omega(t+r_*)}, \label{eq:timofeev}
\end{align}
for some constants $\zeta_\pm$ and functions $\xi(r)$, $\varphi_\pm(r)$, with $r_*$ defined in Eq.~\eqref{def:tortoise}. These geometries present solutions that are manifestly ingoing at the horizon, with $\ell_-^a$ being the degenerate PND, as they fulfil the condition of Eq.~\eqref{eq:perturbed_ingoing}. Given some action, one can then use this ansatz in the  equations of motion of the theory. In general, the ansatz will solve the equations of motion only for specific frequencies $\omega = \omega_*$, namely, the  algebraically special frequencies. We also note that this ansatz is highly reminiscent of the original metric ansatz used to uncover the phenomenon of pole skipping \cite{Grozdanov:2017ajz}. 

A number of additional comments about the ansatz \eqref{eq:timofeev} are in order. Firstly, and perhaps obviously, $\xi(r)$, $\varphi_+(r)$ and $\zeta_+$ are in the even channel of perturbations whereas $\varphi_-(r)$ and $\zeta_-$ are in the odd channel. Generally, the channels will not be non-zero simultaneously for a given algebraically special solution. The even channel of the ansatz obeys the shear-free condition \eqref{eq:shear-free} in the linearised sense. Furthermore, setting $\varphi_\pm(r)=\zeta_-=0$, one obtains a linearised version of the Robinson-Trautman spacetime \cite{Robinson:1962zz} (see also Refs.~\cite{Bakas:2014kfa, Skenderis:2017dnh}), which, by definition, admits a shear-free, twist-free geodesic null congruence. These geometries (within the even channel) are exhaustive for all the cases discussed in this section, and define, as a matter of convention, the algebraically special frequency $\omega_*$. The odd channel algebraically special solutions are acquired by turning on $\varphi_-(r)$ and $\zeta_-$. In all examples considered in this paper, the odd channel ansatz solves the equations of motion precisely when $\omega=-\omega_*$. We are unaware of a fundamental reason why the algebraically special frequencies in the two channels are the negatives of each other, and to what extent this statement is universal. It would be interesting to see potential counterexamples and understand the underlying geometric reasons behind either of the two possible scenarios.

\section{The structure of dualities}
\label{sec:dualities}
It is well known that the classical vacuum Maxwell theory is invariant under the exchange of the electric and magnetic field, known in some contexts as the S-duality. This is a consequence of the fact that the equations of motion and Bianchi identities exhibit the same form and is formally expressed by swapping the field strength tensor $F_{ab}$ with its Hodge dual $\widetilde{F}_{ab}$, at least locally. For example, on the background of an electrically charged black hole, such a transformation would yield an electrically neutral black hole with finite magnetic charge.

On sufficiently simple backgrounds, the linearised metric exhibits a similar structure \cite{Hull:2001iu,Henneaux:2004jw,Julia:2005ze,Nieto:1999pn,deHaro:2008gp,Leigh:2003ez}, where the Weyl tensor $C_{abcd}$ and its dual $\widetilde C_{abcd}$ are swapped. Non-linear extensions have also been studied (see e.g.~Ref.~\cite{Pereniguez:2023wxf,Kol:2020ucd,Huang:2019cja,Kol:2019nkc,Emond:2020lwi}). In this paper, however, we will work with the duality transformations that only affect the perturbed part of the fields, leaving the black hole background intact. More precisely, we study the duality maps that map solutions from the odd channel of perturbations to the even channel, and vice versa \cite{Chandrasekhar:1976zz,Chandrasekhar:1975zza,Chandrasekhar:1979iz,Chandrasekhar:1984mgh,Chandrasekhar:1985kt,Anderson:1991kx,heading:1997,chandrasekhar:1980a,Glampedakis:2017rar,Lenzi:2021njy,Bakas:2014kfa,Yurov:2018ynn,Bakas:2008gz,Grozdanov:2023txs}. Here, we build a firmer geometric understanding of these maps, and show how they reduce to the common electric-magnetic duality on trivial backgrounds, be it for the metric tensor or the Maxwell field. Specifically, we focus on the Einstein-Maxwell action and background geometries from Eq.~\eqref{def:background_metric}.

\subsection{Equations of motion}
Gravitational perturbations around maximally-symmetric black hole backgrounds are most commonly treated by combining the perturbed fields into gauge-invariant master variables $\psi(r)$ that reduce the perturbed Einstein's equations into a set of \emph{master equations} in the Schr\"{o}dinger-like form (see e.g.~Ref.~\cite{Chandrasekhar:1985kt,Kodama:2003jz,Kodama:2003kk,Martel:2005ir}):
\begin{equation}
    \qty(\frac{d^2}{dr_*^2}+\omega^2-V(r))\psi(r)=0. \label{eq:master}
\end{equation}
This approach is appealing due to its analogy with the one-dimensional scattering theory. However, it is also inconvenient for the following two reasons: Firstly, the master functions $\psi(r)$ have no direct geometric or physical interpretation, and, secondly, they depend linearly both on the perturbed fields and their radial derivatives. This means that, while the ingoing/outgoing boundary conditions at horizons or asymptotically flat regions of spacetimes translate directly from the fields $\delta g(r)$ and $\delta A(r)$ to the master functions $\psi(r)$, the Dirichlet or Neumann boundary conditions do not. Consequently, the master functions $\psi(r)$ may not be most convenient for AdS/CFT calculation that use the standard holographic dictionary.

\subsection{Darboux transformations}
\label{subsec:Darboux}
In terms of the master equations \eqref{eq:master}, the duality transformations manifest themselves in the form of {\em Darboux pairs} \cite{Glampedakis:2017rar,Lenzi:2021njy,Grozdanov:2023txs}---a structure analogous to the one known from supersymmetric quantum mechanics \cite{Cooper:1994eh,Liu:1997qn,Gu:2005,Anninos:2011af}. A pair of equations for two decoupled variables $\psi_+$ and $\psi_-$ (in this context, they are to be understood as a pair of decoupled even/odd variables),
\begin{subequations}
\label{eq:evenodd}
\begin{align}
    \qty(\frac{d^2}{dr_*^2}+\omega^2-V_+(r))\psi_+(r)&=0,\label{eq:even}\\
    \qty(\frac{d^2}{dr_*^2}+\omega^2-V_-(r))\psi_-(r)&=0,\label{eq:odd}
\end{align}
\end{subequations}
can admit a structure that allows us to express a solution of Eq.~\eqref{eq:even} using a solution of Eq.~\eqref{eq:odd} in a simple manner by using a first-order differential operator. This is possible if we can express the potentials $V_\pm(r)$ as
\begin{equation}
    V_\pm(r)=\omega_*^2+L_\pm W(r), \label{def:DarbouxPotentials}
\end{equation}
for some function $W(r)$ and a constant $\omega_*$. The differential operators $L_\pm$ are defined as
\begin{equation}
    L_\pm=W(r)\pm \frac{d}{dr_*}. \label{def:Lpm}
\end{equation}
We can then express Eqs.~\eqref{eq:evenodd} in the following way:
\begin{subequations}
\label{eq:evenodd1}
\begin{align}
    L_+ L_- \psi_+(r)&=(\omega^2-\omega_*^2)\psi_+(r), \label{eq:even1}\\
    L_- L_+ \psi_-(r)&=(\omega^2-\omega_*^2)\psi_-(r). \label{eq:odd1}
\end{align}
\end{subequations}
This immediately implies that, given a solution  $\psi^1_+(r)$ of Eq.~\eqref{eq:even1}, the function
\begin{subequations}
\label{eq:dualitymaps}
\begin{equation}
\label{eq:even2odd}
    \psi^2_-(r)=L_- \psi^1_+(r)
\end{equation}
solves Eq.~\eqref{eq:odd1}, and given a solution $\psi^1_-(r)$ of Eq.~\eqref{eq:odd1}, the function
\begin{equation}
\label{eq:odd2even}
    \psi^2_+(r)=L_+ \psi^1_-(r)
\end{equation}
\end{subequations}
solves Eq.~\eqref{eq:even1}. Here, we have introduced superscripts $1,2$ to emphasise that Eqs.~\eqref{eq:dualitymaps} are maps in the space of solutions rather than on-shell conditions. Such maps are known as {\em Darboux transformations}. They provide the relevant maps between the even and odd channels and are invertible for $\omega^2 \neq \omega^2_*$. To see this, consider setting $\omega^2=\omega_*^2$. Eqs.~\eqref{eq:evenodd1} can then solved by solving the first-order equations
\begin{subequations}
\label{eq:annihilators}
\begin{align}
    L_- \psi_+(r)&=0, \\
    L_+ \psi_-(r)&=0. 
\end{align}
\end{subequations}
The solutions to these equations are
\begin{subequations}
\label{def:algspec}
\begin{align}
    \psi_+(r)=\qty[\phi(r) e^{i \omega_* r_*}]^{-1},\\
    \psi_-(r)=\qty[\phi(r) e^{i \omega_* r_*}]^{+1},
\end{align}
\end{subequations}
where $\phi(r)$ is a function related to $W(r)$ via
\begin{equation}
    W(r)=-\frac{d}{dr_*}\ln\phi(r)-i\omega_*. \label{def:W}
\end{equation}
Note that the potentials $V_\pm(r)$ and, therefore, the master equations in both channels are completely determined by $f(r)$, $\phi(r)$ and the constant $\omega_*$. As we explain in the next section (see also Refs.~\cite{Chandrasekhar:1984mgh,Bakas:2014kfa}), it turns out that the solutions \eqref{def:algspec} correspond to the algebraically special solutions of Section~\ref{subsec:algebraically_linear}, with $\pm\omega_*$ the corresponding algebraically special frequencies, where $\omega^2=\omega_*^2$ is a necessary condition for Eqs.~\eqref{eq:perturbed_conditions} to hold.

Whenever the ingoing or the outgoing interpretation of $\psi(r)$ is applicable (see Section~\ref{subsec:gravitational_waves}), the operators $L_\pm$ map the ingoing waves to the ingoing waves, and the outgoing waves to the outgoing waves. In other words, the duality maps \eqref{eq:dualitymaps} preserve the ingoing/outgoing nature of a solution around the event/cosmological horizon or at asymptotic infinity. In asymptotically flat or asymptotically de Sitter spaces, this results in the isospectrality of quasinormal modes (defined by being ingoing for $r_* \rightarrow -\infty$ and outgoing for $r_* \rightarrow \infty$; see e.g.~Ref.~\cite{Konoplya:2011qq}) between the channels. In asymptotically anti-de Sitter spaces, quasinormal modes are defined by the Dirichlet or mixed boundary conditions, which are not preserved by the operators $L_\pm$. Isospectrality is thus replaced by a much more subtle structure, namely, the \emph{spectral duality relation} constructed in our Ref.~\cite{Grozdanov:2024wgo} and discussed at length in Section~\ref{sec:AdSCFT}.

\subsection{The self-dual limit}
\label{sec:self-dual}
Next, we comment on the $\omega_*\rightarrow \infty$ limit of the above construction. Suppose that, by varying some parameter $\eta$, we have
\begin{align}
    \eta&\rightarrow 0:& \omega_* &\sim \frac{\omega_*^{(0)}}{\eta} ,\\
    & & \phi(r) &\sim \phi^{(0)}+\phi^{(1)}(r)\eta  .\nonumber  
\end{align}
The limit $\eta\rightarrow 0$ of Eqs.~\eqref{eq:evenodd1} is then well defined and yields the \emph{same} master equations of the Schr\"odinger form \eqref{eq:master} in both channels. The corresponding potentials are
\begin{equation}
    \eta\rightarrow0:\qquad V_+(r)=V_-(r)=\frac{2i \omega_*^\qty(0)}{\phi^\qty(0)} \frac{d\phi^\qty(1)(r)}{dr_*}.
\end{equation}
In this limit, there are no algebraically special solutions in the sense of Eqs.~\eqref{def:algspec}, as the corresponding algebraically special frequencies are `pushed to infinity'.  As in Ref.~\cite{Grozdanov:2024wgo}, we refer to this limit as the \emph{self-dual limit}. In terms of the master functions $\psi_\pm(r)$, this limit means that both functions obey the same master equation \eqref{eq:master}. It is this self-dual limit that reduces to the familiar cases of the electric-magnetic duality. The analogue of the duality maps \eqref{eq:dualitymaps} is then simply the identity map. Furthermore, note that a Darboux pair of potentials $V_+(r)$ and $V_-(r)$ can obey $V_+(r)=V_-(r)$ only in the sense of a limit described above, or if the potentials are constant, $V_+(r)=V_-(r)=\text{const.}$ It is therefore appropriate to look at two variables obeying the same master equation in terms of a limiting (self-dual) case of a certain Darboux pair. 

\subsection{Examples}
\label{subsec:examples}
\subsubsection{The Schwarzschild black hole}
\label{subsubsec:schw}
As our first example, we consider the metric and Maxwell perturbations on the background  of an uncharged Schwarzschild black hole with an arbitrary cosmological constant. In this case, the metric and the Maxwell field perturbations are completely decoupled at the linear level. Moreover, due to the symmetry of the problem, the analysis splits into decoupled even $+$ and odd $-$ channels. Specifically, we are solving
\begin{subequations}
\label{eq:EinsteinMaxwell}
\begin{align}
    R_{ab}=\Lambda g_{ab},\\
    \nabla_a F^{ab}=0.
\end{align}
\end{subequations}
The background metric is given by Eq.~\eqref{def:background_metric}, with
\begin{equation}
    f(r)=K-\frac{2M}{r}-\frac{\Lambda}{3}r^2,
\end{equation}
where $M$ is the mass of the black hole and $\Lambda$ the cosmological constant. The dynamics of the Maxwell field is governed by the master equation \eqref{eq:master}, with the same potential in both the even and the odd channels:
\begin{equation}
\label{eq:VMaxwell}
    V_\pm^\text{Maxwell}=\mu\frac{f(r)}{r^2},
\end{equation}
with the appropriate master functions defined in Appendix~\ref{app:master}. In Section~\ref{subsub:RN}, we will see how the coinciding potentials arise as a consequence of the self-dual, zero-charge limit of the Reissner-Nordstr\" om black hole. The even and odd master functions $\psi_\pm(r)$ (see Appendix~\ref{app:master} for definitions) form a Darboux pair with \cite{Glampedakis:2017rar,Lenzi:2021njy}
\begin{align}
    \omega_*&=i\frac{\mu(\mu-2K)}{12M},\label{def:AS}\\
    \phi(r)&=\mu-2K+\frac{6M}{r}.
\end{align}
This can be checked explicitly using the potentials from, e.g.,  Refs.~\cite{Kodama:2003jz,Kodama:2003kk,Martel:2005ir}. Instead, here, we attempt to shed some light on the physical reasons behind the emergence of the Darboux structure, which is completely obscured by the master function formalism.

We first consider the Maxwell perturbations. In terms of the electric and magnetic variables introduced in Eqs.~\eqref{def:CECB}, we can express the Maxwell equations in the even channel as
\begin{subequations}
\label{eq:evenMaxwell}
    \begin{align}
        \frac{\CE_+}{\sqrt{f(r)}}&=\frac{i\omega r^2}{r^2\omega^2-\mu f(r)}\partial_r\qty({\sqrt{f(r)}\CB_+}),\\
        \frac{\CB_+}{\sqrt{f(r)}}&=-\frac{1}{i\omega}\partial_r\qty({\sqrt{f(r)}}\CE_+).
    \end{align}
\end{subequations}
The equations are invariant with respect to $(\omega,\CE_+)\rightarrow(-\omega,-\CE_+)$ and $(\omega,\CB_+)\rightarrow(-\omega,-\CB_+)$. Consequently, the second-order equation of motion for the ingoing wave ($\CE_+-\CB_+$) is the time-reversed version of the equation of motion for the outgoing wave ($\CE_++\CB_+$). Explicitly,
\begin{subequations}
\label{eq:maxwellEven}
\begin{align}
    \CF(\omega,r)\qty[r \CE_+ + r \CB_+]&=0, \label{eqs:mx1}\\
    \CF(-\omega,r)\qty[r \CE_+ - r \CB_+]&=0, \label{eq:mx2}
\end{align}
\end{subequations}
where we have defined a differential operator
\begin{equation}
\label{def:CF}
    \CF(\omega,r)=\frac{d^2}{dr_*^2}+\omega^2-\alpha(\omega,r),
\end{equation}
with
\begin{align}
\label{def:alpha}
\alpha(\omega,r)&=\frac{f(r)}{r^2}\qty[\kappa+\mu-2ir\omega-f(r)]\nonumber\\&+\frac{1}{4}f'(r)\qty[4i\omega+f'(r)]. 
\end{align}
Either of the Eqs.~\eqref{eq:maxwellEven} can be used as an appropriate master equation, with any given boundary condition, encoding the physics of the scattering process at hand. The equations \eqref{eq:maxwellEven} are not independent, i.e., there is only one specific solution of Eq.~\eqref{eq:mx2} that corresponds to any given solution of its time-reversed counterpart \eqref{eqs:mx1}, as dictated by Eqs.~\eqref{eq:evenMaxwell}. The situation in the odd channel is completely analogous, with the Maxwell equations being
\begin{subequations}
    \begin{align}
        \frac{\CE_-}{\sqrt{f(r)}}&=\frac{1}{i\omega}\partial_r\qty({\sqrt{f(r)}}{\CB_-}),\\
        \frac{\CB_-}{\sqrt{f(r)}}&=-\frac{i\omega r^2}{r^2\omega^2-\mu f(r)}\partial_r\qty({\sqrt{f(r)}}{\CE_-}),
    \end{align}
\end{subequations}
and the corresponding pair of second-order equations of motion for the ingoing and outgoing waves given by the following expressions:
\begin{subequations}
\begin{align}
    \CF(\omega,r)\qty[r \CE_- - r \CB_-]&=0,\\
    \CF(-\omega,r)\qty[r \CE_- + r \CB_-]&=0.
\end{align}
\end{subequations}
Evidently, the equation for the ingoing/outgoing wave in one channel is the same as the equation for the ingoing/outgoing wave in the other channel.

The manifestation of the electric-magnetic duality is now clear. Given a solution $\CE^1_+(r)$, $\CB^1_+(r)$ in the even channel, we can obtain a solution $\CE^2_+(r)$, $\CB^2_+(r)$ in the odd channel (up to a multiplicative constant) via
\begin{subequations}
\label{eq:dualityEM}
\begin{align}
    \CE^2_-+\CB^2_- = +\qty(\CE^1_+-\CB^1_+), \\
    \CE^2_--\CB^2_- = -\qty(\CE^1_++\CB^1_+), 
\end{align}
where we have introduced a superscript to indicate that the duality transformations \eqref{eq:dualityEM} are not on-shell conditions, but rather maps in the space of solutions. Analogously, we can map a solution of the odd channel to a solution of the even channel as
\begin{align}
    \CE^2_++\CB^2_+&=-\qty( \CE^1_--\CB^1_-), \\
    \CE^2_+-\CB^2_+&=+\qty(\CE^1_-+\CB^1_-).
\end{align}
\end{subequations}
Eqs.~\eqref{eq:dualityEM} express the standard electric-magnetic duality $\vb{E}\rightarrow \vb{B}$, $\vb{B}\rightarrow-\vb{E}$, written in a way that can be easily translated into the treatment of gravitational waves.

Extending the discussion above to the dynamics of the perturbed metric, we introduce, analogously to Eq.~\eqref{def:CF},
\begin{equation}
    F(\omega,r)=\frac{d^2}{dr_*^2}+\omega^2-a(\omega,r) \label{def:F}
\end{equation}
with
\begin{align}
    a(\omega,r)&=\frac{f(r)}{r^2}\qty[\mu+2K+4ir \omega-r f'(r)-2 f(r)]\nonumber\\&+f'(r)\qty[f'(r)-2i\omega].
\end{align}
Using the gravitational variables introduced in Eqs.~\eqref{def:EB}, we can write the even channel perturbation equations as (cf.~Eq.~\eqref{eq:maxwellEven} for the Maxwell field)
\begin{subequations}
\label{eq:evenEoM}
\begin{align}
    F(\omega,r)\qty[r E_+ + r B_+]&=0,\label{eq:Fexample}\\
    F(-\omega,r)\qty[r E_+ - r B_+]&=0.
\end{align}
\end{subequations}
The equations for the ingoing and outgoing perturbations are time-reversed versions of one another, and have to be solved simultaneously. The constraint equations relating $E_+$ and $B_+$ are now much more complicated, and are presented in Appendix~\ref{app:master}. The odd channel is, again, completely analogous to the Maxwell case, obeying the same set of equations as the even channel, namely, 
\begin{subequations}
\label{eq:oddEoM}
\begin{align}
    F(\omega,r)\qty[r E_- - r B_-]&=0,\\
    F(-\omega,r)\qty[r E_- + r B_-]&=0.
\end{align}
\end{subequations}
The equations enjoy a discrete invariance, precisely reflecting Darboux transformation of Section \ref{subsec:Darboux}. This invariance is analogous to the electric-magnetic invariance of the Maxwell case, but this time with additional algebraically special structure. Explicitly, we can write down the Darboux map \eqref{eq:even2odd}, mapping an even solution $E_+^1$, $B_+^1$ to an odd solution $E_-^2, B_-^2$, as
\begin{subequations}
\label{eq:dualityEMmetric}
\begin{align}
E^2_-+B^2_-&=-i(\omega-\omega_*)(E^1_+-B^1_+),\\
E^2_--B^2_-&=-i(\omega+\omega_*)(E^1_++B^1_+).\label{eq:example}
\end{align}
Correspondingly, using the Darboux map $\eqref{eq:odd2even}$, we can map an odd solution $E_-^1$, $B_-^1$ to an even solution $E_+^2$, $B_+^2$, 
\begin{align}
E^2_++B^2_+&= +i(\omega-\omega_*)(E^1_--B^1_-),\\
E^2_+-B^2_+&= +i(\omega+\omega_*)(E^1_-+B^1_-).
\end{align}
\end{subequations}
The scaling factors in Eqs.~\eqref{eq:dualityEMmetric} are chosen so that all variables remain finite for all choices of $\omega$. 

The duality maps in the form of Eqs.~\eqref{eq:dualityEMmetric} offer several insights. Firstly, they parallel the electric-magnetic duality of Maxwell equations, which, as noted above, is difficult to see in the language of master functions and Darboux transformations. Unlike in the Maxwell case, however, the maps now mix the electric and magnetic variables. In the self-dual limit, when $M\rightarrow 0 $ and $\omega_* \rightarrow \infty$, the duality maps reduce to the well-known electric-magnetic duality for the metric on the backgrounds of maximally symmetric spaces \cite{Hull:2001iu,Henneaux:2004jw,Julia:2005ze,Nieto:1999pn,deHaro:2008gp,Leigh:2003ez}. In terms of the master functions, this limit yields the same potentials for both channels, as was mentioned in Section~\ref{sec:self-dual}, namely,
\begin{equation}
    V_\pm^\text{metric}=\mu\frac{f(r)}{r^2}.
\end{equation}
Secondly, the duality maps of Eqs.~\eqref{eq:dualityEMmetric} are simple transformations that do not have any radial dependence or derivatives. This is in contrast with the Darboux maps, which are expressed in terms of the $r$-dependent first-order linear operators, resulting from the master function formalism and the demand for the potential to be $\omega$-independent. Thirdly, the duality maps of Eqs.~\eqref{eq:dualityEMmetric} make the non-invertibility at $\omega^2=\omega_*^2$ transparent, since the variables onto which the duality acts naturally are precisely the variables that vanish for the algebraically special solutions. As in the Maxwell case, any one of Eqs.~\eqref{eq:evenEoM} can be used as an even channel master equation, and any of Eqs.~\eqref{eq:oddEoM} can be used as an odd channel master equation, as long as $\omega^2\neq \omega_*^2$. This is due to fact that, at $\omega^2=\omega_*^2$, the maps between the relevant variables can be singular, as can be inferred from the discussion of Section~\ref{subsec:algebraically_linear}. This is made more transparent in Appendix~\ref{app:master}.

\subsubsection{The Reissner-Nordstr\"om black hole}
\label{subsub:RN}
Next, we consider the coupled Einstein-Maxwell equations
\begin{subequations}
\begin{align}
    R_{ab}&=\Lambda g_{ab}+\kappa^2 \qty(T_{ab}-\frac{1}{2}T^c{}_c g_{ab}),\\
    \nabla_a F^{ab}&=0,
\end{align}
\end{subequations}
with the energy-momentum tensor given by
\begin{equation}
    T_{ab}=-\frac{1}{4}F^{cd}F_{cd} g_{ab}-F_{ac}F^c{}_d.
\end{equation}
We are interested in the solutions of the Reissner-Nordstr\"om type with
\begin{subequations}
\begin{align}
    f(r)&=K-\frac{2M}{r}-\frac{\Lambda}{3}r^2+\frac{Q_e^2}{r^2},\\
    A_a dx^a&=\frac{\sqrt{2} Q_e}{\kappa r} dt.
\end{align}
\end{subequations}
Here, $Q_e$ is the electric charge and $\kappa$ is the coupling constant between gravity and the Maxwell field. For convenience, we define
\begin{subequations}
\begin{align}
    Q&=\frac{Q_e}{3M},\\
    \Delta^\qty(\pm)&=\frac{1}{2}\qty[1\pm\sqrt{1+4Q^2(\mu-2K)}].
\end{align}
\end{subequations}
In both the even and the odd channel of perturbations, the analysis splits into two independent sectors of mixed metric-Maxwell perturbations, denoted by the superscripts $(+)$ and $(-)$. The four independent equations of motion are organised into two Darboux pairs. In the $(+)$ sector, we have
\begin{subequations}
\begin{align}
    L_+^\qty(+)L_-^\qty(+) \psi_+^\qty(+)(r)&=\qty(\omega^2-\omega_\qty(+)^2)\psi_+^\qty(+)(r),\\
    L_-^\qty(+)L_+^\qty(+) \psi_-^\qty(+)(r)&=\qty(\omega^2-\omega_\qty(+)^2)\psi_-^\qty(+)(r).
\end{align}
\end{subequations}
In the $(-)$ sector, we have
\begin{subequations}
\begin{align}
    L_+^\qty(-)L_-^\qty(-) \psi_+^\qty(-)(r)&=\qty(\omega^2-\omega_\qty(-)^2)\psi_+^\qty(-)(r),\\
    L_-^\qty(-)L_+^\qty(-) \psi_-^\qty(-)(r)&=\qty(\omega^2-\omega_\qty(-)^2)\psi_-^\qty(-)(r).
\end{align}
\end{subequations}
The operators $L^\qty(\pm)_\pm$ are defined through Eqs.~\eqref{def:Lpm} and \eqref{def:W} with
\begin{align}
    \omega_\qty(\pm)&=\frac{\omega_*}{\Delta^\qty(\pm)},\\
    \phi^\qty(\pm)&=\mu-2K+\frac{6M}{r}\Delta^{\qty(\pm)},
\end{align}
where $\omega_*$ is defined in Eq.~\eqref{def:AS}. An explicit calculation shows that the corresponding set of algebraically special solutions (see Eq.~\eqref{def:algspec}) is consistent with the ansatz of Eq.~\eqref{eq:timofeev}, where it should be emphasised that there are now two sets of pairwise opposite algebraically special frequencies where the solution can be algebraically special, namely, $\omega=\pm\omega_\qty(+)$ and $\omega=\pm\omega_\qty(-)$.

In the zero-charge limit $Q_e\rightarrow 0,$ we have
\begin{subequations}
\begin{align}
    Q_e &\rightarrow 0: & \Delta^\qty(+) &\rightarrow 1, \\
    & & \Delta^\qty(-) &\rightarrow 0,
\end{align}
\end{subequations}
and, consequently,
\begin{subequations}
\begin{align}
    Q_e &\rightarrow 0: & \omega_\qty(+) &\rightarrow \omega_*, \\
    & & \omega_\qty(-) &\rightarrow \infty.
\end{align}
\end{subequations}
The $(+)$ sector reduces to purely metric perturbations and the corresponding Darboux pair, whereas the $(-)$ sector exhibits the self-dual behaviour in the sense of Section~\ref{sec:self-dual}. As a consequence, the master potentials for the Maxwell perturbations in this limit coincide (see Eq.~\eqref{eq:VMaxwell}).

For the charged case, we do not present a full analysis in terms of the electric and magnetic variables. Due to the non-trivial dependence of $F_{ab}$ on the linearised diffeomorphisms, it is difficult to find `natural' gauge-invariant variables corresponding to electric- and magnetic-like fields that would make the duality structure more transparent. Nevertheless, Chandrasekhar's original treatment using the Newman-Penrose formalism \cite{Chandrasekhar:1979iz,Chandrasekhar:1985kt} suggests that such variables should exist. 

\subsubsection{The linear axion model}
\label{subsubsec:linear_axion0}
In this section, we turn our attention to a simple model of 4$d$ Einstein gravity with two minimally coupled massless scalar fields (axions), $\Phi_y$ and $\Phi_z$. We consider a planar black hole ($K=0$) in asymptotically anti-de Sitter space, with the scalar fields varying linearly along the boundary coordinates. This is the linear axion model introduced in Ref.~\cite{Andrade:2013gsa}.\footnote{We do not consider the Maxwell field in this section.} For simplicity, we choose $Y=e^{ikz}$ and $\mu=k^2$ in the $(t,r,y,z)$ coordinate system, and rescale the cosmological constant to equal $\Lambda=-3$. The equations of motion are then
\begin{subequations}
\begin{align}
    R_{ab}&=-3g_{ab}+T_{ab}-\frac{1}{2}T^c{}_c g_{ab},\\
    \nabla_a \nabla^a \Phi_{(i)}&=0,
\end{align}
\end{subequations}
where
\begin{subequations}
\begin{align}
    T^{(i)}_{ab}&=-\frac{1}{4}g_{ab}\qty(\partial_a \Phi_{(i)} \partial^a \Phi_{(i)})+\frac{1}{2}\partial_a \Phi_{(i)} \partial_b \Phi_{(i)},\\
    T_{ab}&=T^{(y)}_{ab}+T^{(z)}_{ab}.
\end{align}
\end{subequations}
The solutions of interest are
\begin{subequations}
    \begin{align}
    f(r)&=-\frac{2M}{r}-\frac{m^2}{2}+r^2 ,\\
    \Phi_y&=my,\\
    \Phi_z&=mz.
\end{align}
\end{subequations}
Note that the background retains the $\partial_y$ and $\partial_z$ isometries even in the presence of the translational-symmetry-breaking parameter $m$.

As in the previous cases, we perturb the metric, as well as the scalars, $\Phi_{(i)}\rightarrow\Phi_{(i)}+\delta \Phi_{(i)}$. The theory has four independent channels of perturbations, and four corresponding master functions (see Appendix~\ref{app:master} for definitions) obeying the master equation \eqref{eq:master} for various potentials. The channels that reduce to purely scalar perturbations in the $m\rightarrow 0 $ limit are controlled by the potentials
\begin{equation}
    V_{y,z}=f(r)\frac{2k^2+m^2+6 r^2-2f(r)}{2r^2}.
\end{equation}
The channels that reduce to purely metric perturbations in the $m\rightarrow 0 $ limit form an even/odd Darboux pair, controlled by
\begin{subequations}
    \begin{align}
    \omega_*&=i\frac{k^4+k^2 m^2}{12M},\\
    \phi(r)&=k^2+m^2+\frac{6M}{r}.
    \end{align}
\end{subequations}
The corresponding algebraically special solutions are consistent with the ansatz of Eq.~\eqref{eq:timofeev}.

The self-dual limit of $M\rightarrow 0 $ is particularly interesting in this model \cite{Davison:2014lua}. In this limit, the algebraically special frequency $\omega_*$ diverges, yielding the same master equations for both the even and the odd channels. The background geometry retains an event horizon and becomes conformal to a patch of AdS$_2 \times \mathbb{R}^2$, with the equations of motion enjoying an $\text{SL}(2,\mathbb{R}) \times \text{SL}(2,\mathbb{R})$ symmetry. This allows for a closed-form solution of all the perturbation equations to be given in terms of hypergeometric functions. This self-dual limit was discussed in detail in Ref.~\cite{Davison:2014lua} (see also Section~\ref{subsubsec:axion}), and we do not repeat the analysis here. It has been noted \cite{Andrade:2013gsa} that the linearised linear axion model shares many quantitative features with a massive gravity model \cite{Vegh:2013sk,Davison:2013jba,Blake:2013bqa}. We did not investigate the duality relations in theories of massive gravity, but we have verified that the algebraically special frequencies coincide with those of the linear axion model.

\subsection{Pole skipping}
The equations of motion for perturbations around a background in two-derivative theories are second order differential equations and therefore have two solutions. At a horizon (be it the event horizon or the cosmological horizon), one generally imposes either the ingoing or the outgoing boundary conditions. Normally, this uniquely (up to a scaling constant) determines the solution of the equations of motion at the horizon. Pole skipping \cite{Grozdanov:2017ajz,Grozdanov:2019uhi,Blake:2018leo,Blake:2017ris,Blake:2019otz} (see also \cite{Grozdanov:2018kkt,Grozdanov:2020koi,Wang:2022mcq,Natsuume:2019xcy,Ahn:2020baf,Blake:2021hjj,Ceplak:2019ymw,Abbasi:2020xli,Grozdanov:2023tag,Choi:2020tdj,Ning:2023ggs,Chua:2025vig}) is a phenomenon in which such a solution, however, fails to be unique. That is, we still retain two linearly independent ingoing solutions. This can happen at the $(\Omega_n, \mu_n)$ points, where $\Omega_n$ is a multiple of the imaginary Matsubara frequency
\begin{equation}
    \Omega_n=-2\pi i n T,
\end{equation}
with $T$ being the Hawking temperature of the black hole.\footnote{Care should be taken when considering the cosmological horizon, where the temperature is defined with a minus sign \cite{Spradlin:2001pw,Grozdanov:2023txs}.} The values of $\mu_n$ at which the pole-skipping phenomenon exists may be complex.

It was shown in Ref.~\cite{Grozdanov:2023txs} that the pole-skipping points of the metric perturbations in the 4$d$ Schwarzschild case split into an infinite set of pole-skipping points that are common between both channels, and the algebraically special pole-skipping points for which (see also Ref.~\cite{Grozdanov:2020koi})
\begin{equation}
    \Omega_n^2=\omega_*^2 .
\end{equation}
The same discussion can be immediately generalised to all scenarios, which can be formulated in terms of Darboux pairs discussed in Section~\ref{subsec:examples}, as long as the function $\phi(r)$ is analytic and finite at the horizon. If we consider the pole-skipping points associated with the ingoing boundary conditions at the outer event horizon of a black hole, this implies a set of even-channel pole-skipping points
\begin{subequations}
\label{eq:pole-skipping}
\begin{equation}
    \Omega_n=\omega_*, \quad n\geq 1,
\end{equation}
and a set of odd-channel pole-skipping points
\begin{equation}
    \Omega_n=-\omega_*, \quad n\geq 1,
\end{equation}
\end{subequations}
with the remaining pole-skipping points being common to both channels. The hydrodynamic pole-skipping points at $n=0$ and the chaotic pole-skipping point at $n=-1$ require special attention and we do not discuss them here (see Ref.~\cite{Grozdanov:2023txs}). We note, however, that in the master function formalism, the chaotic pole-skipping point at $n=-1$ emerges when the zero of the algebraically special solution coincides with the event horizon, i.e., when $\phi(r_0)=0$ (or, in the case of the Reissner-Nordstr\" om black hole, when $\phi^{(+)}(r_0)\phi^{(-)}(r_0)=0$). This follows from the definitions of the even-channel master functions (see Appendix~\ref{app:master}), where $\phi(r)$ appears in the denominator.

The fact that the algebraically special solutions at the Matsubara frequencies exhibit pole skipping is interesting in and of itself. It emerges from the fact that, in theories considered in this paper, the ansatz \eqref{eq:timofeev} does not have any logarithmic terms in its Frobenius expansion around the horizon. While we are not aware to what extent this is universal, we have failed to find explicit counterexamples. It would be interesting to better understand the interplay between pole skipping and algebraically special solutions, both in 4$d$ and in more dimensions \cite{Dias:2013hn,Podolsky:2006du}, building on the linearised version of the null shockwave geometry that originally led to pole skipping and was investigated in Ref.~\cite{Grozdanov:2017ajz}.

\section{Asymptotically anti-de Sitter space and holography}
\label{sec:AdSCFT}

Next, we use the holographic duality to investigate role of the gravitational constraints of 4$d$ bulk theories on their 3$d$ CFT duals without gravity. The black hole geometries of Eq.~\eqref{def:background_metric} are dual to finite temperature states of the CFT, where the temperature of the CFT is given by the Hawking temperature. The dualities of Section~\ref{sec:dualities} translate onto dualities between the retarded two-point correlators in the boundary theory. Applications of such dualities in a holographic setting have been previously studied in Refs.~\cite{Grozdanov:2024wgo,Herzog:2007ij,Bakas:2008gz,Bakas:2014kfa,Grozdanov:2023txs,ioannisTalk}.

The retarded correlators of the boundary theory are (almost) completely determined by the spectrum of their corresponding quasinormal modes. This allows us to cast the duality structure into the spectral duality relation \cite{Grozdanov:2024wgo}, which imposes stringent constraints on the quasinormal modes of the two channels of perturbations. Here, we perform a detailed analysis of the correlators in the thermal state in flat 3$d$ Minkowski space, which is dual to the Schwarzschild black brane. We present a collection of numerical results that demonstrate the power of discussed dualities and the spectral duality relation. We then discuss a state with a non-zero chemical potential dual to a charged black brane, and also the linear axion model. Moreover, we describe the physics of the algebraically special solutions from the point of view of the boundary CFT.

While the majority of the contents of this section is applicable to spherical and hyperbolic boundary geometries, as noted above, we limit ourselves to the flat $K=0$ horizon geometry and adopt the standard set of coordinates $(t,r,y,z)$, with $Y=e^{i k z}$ and $\mu=k^2$, where $y=\chi \cos\phi$ and $z=\chi\sin\phi$. Furthermore, we will work in the radial gauge, where $\delta g_{ra}=\delta A_r=0$ and fix the AdS radius by setting the cosmological constant to $\Lambda=-3$. We will henceforth refer to the even/odd variables as the longitudinal/transverse variables, which is more common in the holographic literature. Note that while, here, we will rely on the language of the linear response theory, the results themselves apply directly to the discussion of the quasinormal modes of asymptotically-AdS black holes and black branes without the need for making any references to the holographic boundary QFT.

\subsection{Linear response theory and meromorphic correlators}
\label{subsec:generalities}
The discussion of the linearised perturbations around the bulk black hole can be directly translated into the language of the linear response theory of the boundary theory in a thermal state. The main objects of interest are the retarded two-point correlators (Green's functions), which describe the linear response of observables to external sources. According to the standard holographic dictionary \cite{Witten:1998qj,Gubser:1998bc,Aharony:1999ti}, imposing the Dirichlet boundary conditions (at the AdS boundary ($r\rightarrow \infty)$) on the bulk metric (or Maxwell field) allows us to source the boundary energy-momentum tensor (or the conserved U$(1)$ current) and use small variations to compute the two-point functions in the thermal state.

Schematically, we define the retarded correlators as
\begin{equation}
    G(t,t')=\frac{\delta \expval{\CO(t)}_J}{\delta J(t')},
\end{equation}
where $\expval{O(t)}$ is a real-time (i.e., causal) thermal expectation value of a (Hermitian) operator $\CO$ in the presence of a source $J$. In terms of the bulk, $\expval{\CO}_J$ is the conjugate momentum of the corresponding dynamical field in the radial evolution through the bulk (see Ref.~\cite{Iqbal:2008by}), whereas $J$ is the field's boundary value. We emphasise the use of the variational definition of the retarded correlator, in contrast with the canonical one (see e.g.~Ref.~\cite{Bellac:2011kqa}). The two definitions can differ by contact terms (see e.g.~Refs.~\cite{GV:2025, Romatschke:2009ng}). Importantly, this leads to the fact that the duality equation, which we use in order to derive the spectral duality relation (see Eq.~\eqref{eq:DUALITY} below) is \emph{not} invariant under the modification of the correlators by different contact terms.

We now list several relevant properties of retarded correlators in $\omega$-space (see e.g.~Refs.~\cite{Bellac:2011kqa,Herzog:2009xv,Kovtun:2012rj}). The property
\begin{equation}
    G(\omega)^*=G(-\omega^*) \label{eq:CC}
\end{equation}
allows us to relate the retarded correlator to its corresponding spectral function $\rho(\omega)$
\begin{equation}
    \rho(\omega)=\frac{G(\omega)-G(-\omega)}{2i}.
\end{equation}
For diagonal correlators (i.e., for correlators of the form $\expval{\CO \CO}$ rather than $\expval{\CO \CO'}$), and real $\omega$, we have the positivity property
\begin{equation}
    \omega \rho(\omega) \geq 0. \label{eq:positivity}
\end{equation}
In a stable and causal theory (see e.g. Ref.~\cite{Gulotta:2010cu}), $G(\omega)$ must be analytic for $\Im \omega > 0$. The large-$\omega$ behaviour of $G(\omega)$ is that of a power-law,
\begin{equation}
    \omega \rightarrow \infty: \quad G(\omega) \sim \omega^m, \label{eq:power_law}
\end{equation}
where the limit is taken in any direction of the complex plane where such a limit exists (see Ref.~\cite{GV:2025} for a detailed discussion of this and related complex analytic matters). Since we are working with massless fields, $m$ will generally be an integer. Importantly, in thermal large-$N$ theories with holographic duals, generically, $G(\omega)$ have no branch cuts, only poles in the complex $\omega$ plane (see Refs.~\cite{Hartnoll:2005ju,Kovtun:2005ev,Grozdanov:2016vgg,Grozdanov:2018gfx,Casalderrey-Solana:2018rle,Dodelson:2023vrw,Dodelson:2024atp}). We denote the (typically infinite) set of poles by $\omega_n$, so that 
\begin{equation}
    G(\omega_n)=\infty.
\end{equation}
They correspond to the quasinormal modes of the bulk fields, defined by the appropriate Dirichlet boundary conditions. In other words, $G(\omega)$ are meromorphic functions of $\omega$, and have an infinite number of poles in the lower complex half-plane for real $k$. Eq.~\eqref{eq:CC} then tells us that unless $\omega_n$ is imaginary, the poles must appear in pairs $(\omega_n, -\omega_n^*)$. In this paper, we assume simple (degree-$1$) poles. The final property we make use of is the \emph{thermal product formula} of Ref.~\cite{Dodelson:2023vrw}:
\begin{equation}
\rho(\omega)=\frac{\mu\sinh\frac{\omega}{2T}}{\prod\limits_n \qty(1-\frac{\omega^2}{\omega_n^2})}, \label{def:TPF}
\end{equation}
which was shown to hold for correlators dual to simple bulk wave equations. Here, $\mu$ is a positive momentum- and state-dependent function, which is, most importantly, independent of $\omega$. The product runs over all the QNMs, including all of the mirrored (paired) modes. A residue of a pole $\omega_n$ in $\rho(\omega)$ is then
\begin{equation}
    r_n = -\mu \frac{\omega_n \sinh\frac{\omega_n}{2T}}{2 \prod\limits_{m \neq n}\qty(1-\frac{\omega_n^2}{\omega_m^2})}. \label{eq:residues}
\end{equation}
Knowing the poles (the QNMs) $\omega_n$ and their residues $r_n$ is sufficient for determining the retarded correlator up to a finite number of constants, namely, the contact terms (see Section~\ref{subsec:neutral} and Ref.~\cite{GV:2025}).

In theories considered here, the $3$-dimensional metric on which the boundary theory resides is the regularised boundary value of the bulk induced metric (see Eq.~\eqref{def:induced_metric}):
\begin{equation}
    g^{(0)}_{ab}=\lim_{r\rightarrow\infty}\frac{1}{r^2}h_{ab}(r), 
\end{equation}
and for the gauge field, 
\begin{equation}
    {A}^{(0)}_a=\lim_{r\rightarrow\infty}A_a(r).
\end{equation}
The boundary Levi-Civita tensor is given by 
\begin{equation}
    \epsilon_{abc}=\lim_{r\rightarrow\infty}\frac{1}{r^3}\epsilon_{abcd}n^d,
\end{equation}
and the boundary $\Delta_{ab}$ is given by
\begin{equation}
    \Delta^{(0)}_{ab}=\lim_{r\rightarrow \infty}\frac{1}{r^2}\Delta_{ab}(r). \label{eq:boundary_proj}
\end{equation}
Both $g^{(0)}_{ab}$ and $A^{(0)}_a$ are understood as external sources of the boundary QFT given by the Dirichlet boundary conditions in AdS. Finally, the boundary harmonics simplify to
\begin{subequations}
\begin{align}
    Y_A dx^A&=ik e^{ikz} dz,\\
    X_A dx^A &=ik e^{ikz} dy,\\
    Y_{AB} dx^A  dx^B&=\frac{1}{2}k^2 e^{ikz}(dy^2 - dz^2),\\
    X_{AB} dx^A  dx^B&=-k^2 e^{ikz}dy dz.
\end{align}
\end{subequations}

\subsection{General structure of dualities and the spectral duality relation}
As mentioned in Section \ref{sec:dualities}, the dualities discussed in this paper act only on the linearised perturbations, leaving the background fields intact. Holographically, this is reflected in the fact that the dualities provide a map between the correlators evaluated in the same state. Given that the ingoing behaviour at the horizon is preserved by the generic Darboux maps discussed in Section~\ref{subsec:Darboux}, the response functions of interest are the retarded (or, equivalently, advanced) two point correlators \cite{Son:2002sd}.

The number of independent channels of perturbations in the bulk corresponds to the number of independent retarded correlators in the boundary theory. Much like the full set of perturbed equations of motions can be used to reconstruct all the relevant bulk fields from their respective master equations, the Ward identities on the boundary can be used to relate various CFT correlators (see Section~\ref{subsec:neutral}). Suppose we have a longitudinal/transverse pair of bulk variables that form a Darboux pair. On the boundary, such a duality will generically relate some longitudinal and transverse retarded correlators, $G_+$ and $G_-$, respectively, resulting in the duality equation
\begin{equation}
    G_+(\omega,k) G_-(\omega,k)=\frac{\omega^2}{\omega_*^2(k)}-1. \label{eq:DUALITY}
\end{equation}
Here, $\omega_*$ is the algebraically special frequency of the corresponding Darboux pair, which satisfies the following conditions:
\begin{subequations}
\label{eq:Green_AS}
   \begin{align}
    G_+(\omega_*)&=0,\\
    G_-(-\omega_*)&=0.
\end{align} 
\end{subequations}
The duality equation \eqref{eq:DUALITY} reflects the fact that the Darboux maps, defined by differential operators $L_\pm$, generically fail to preserve the Dirichlet boundary conditions, together with the fact that they are non-invertible for $\omega^2=\omega_*^2$. While the exact definitions of $G_\pm(\omega)$ depend on the specific case at hand, the duality equation implies several robust and generic features that will be described in this section.

The self-dual $\omega_* \rightarrow \infty$ limit of Section \ref{sec:self-dual} reduces to a more familiar form of the duality on the boundary:
\begin{equation}
    G_+(\omega,k) G_-(\omega,k)=-1. \label{eq:SELF_DUALITY}
\end{equation}
Eq.~\eqref{eq:SELF_DUALITY} has a clear interpretation in terms of the boundary CFT as a self-duality of the boundary theory and its particle-vortex dual \cite{Witten:2003ya,Herzog:2007ij}, or its spin-$2$ analogue \cite{Leigh:2003ez,deHaro:2008gp} (see Section \ref{subsec:neutral} for details). This is because Eq.~\eqref{eq:SELF_DUALITY} is a statement about the Dirichlet boundary conditions (standard quantisation) giving the same (two-point) correlators as the Neumann boundary conditions (alternative quantisation), albeit with the two channels swapped. An analogous equation appears also for any large-$N$ CFT, in any number of spacetime dimensions, when relating the `same' correlator in two theories related by the renormalisation group flow due to a double-trace deformations of the CFT \cite{Witten:2001ua,Gubser:2002vv}. We discuss this case extensively in Ref.~\cite{GV:2025}. No such clear interpretation of the more general case \eqref{eq:DUALITY} is presently understood.

Let us now clarify a few formal matters. Firstly, we suppress the explicit dependence on the momentum, meaning that all the constants that appear will be implicitly functions of $k$. Next, we remark that we demand that the correlators $G_\pm(\omega)$ are defined so that they obey the analytic properties described in Section \ref{subsec:generalities}. Importantly, this fixes the ambiguity of the redefinition
\begin{subequations}
    \begin{align}
        G_+(\omega) &\rightarrow \alpha(\omega)G_-(\omega),\\
        G_-(\omega) &\rightarrow \frac{G_-(\omega)}{\alpha(\omega)}.
    \end{align}
\end{subequations}
Here, we demand that $G_\pm(\omega)$ do not have any poles that do not correspond to their respective quasinormal modes. Furthermore, we demand that $G_\pm(0)$ is neither zero nor infinite.\footnote{Note that the correlators here are those of conserved currents, and thus have hydrodynamic poles, which, as $k \to 0$, yield poles at $\omega=0$. We will treat this as a limiting case of the $k\neq 0$ results without encountering any issues.} This constraints $\alpha(\omega)$ to a holomorphic function without zeroes or poles. Finally, the demand for power-law asymptotic behaviour of Eq.~\eqref{eq:power_law} fully constraints $\alpha(\omega)$ to a (possibly $k$-dependent) constant. This is a consequence of the fact that there exist {\em no} non-constant bounded entire functions that would have no zeroes and be asymptotically bounded by a polynomial.

The properties of meromorphic retarded correlators described in Section \ref{subsec:generalities}, together with the duality relation \eqref{eq:DUALITY} and the algebraically special conditions \eqref{eq:Green_AS} are sufficient to express the correlators $G_\pm(\omega)$ in terms of the following infinite products:
\begin{subequations}
\label{eq:Gproducts}
    \begin{align}
    G_+(\omega)=\alpha \qty(\frac{\omega}{\omega_*}-1) \prod_n \frac{1-\frac{\omega}{\omega_n^-}}{1-\frac{\omega}{\omega_n^+}},\\
    G_-(\omega)=\frac{1}{\alpha} \qty(\frac{\omega}{\omega_*}+1) \prod_n \frac{1-\frac{\omega}{\omega_n^+}}{1-\frac{\omega}{\omega_n^-}},
\end{align}
\end{subequations}
where $\qty{\omega_n^+}$ and $\qty{\omega_n^-}$ are the sets of poles in the longitudinal and the transverse channel, respectively, and $\alpha$ is real and independent of $\omega$. To ensure convergence, the infinite products above need to be performed as a limiting procedure that includes all the zeros and poles of the correlator in a disk centred at the origin of the complex plane with radius $R$, taking $R \to \infty$.  

Plugging the product expansions of Eq.~\eqref{eq:Gproducts} into the thermal product formula \eqref{def:TPF} gives the \emph{spectral duality relation} \cite{Grozdanov:2024wgo}:
\begin{equation}
    \label{eq:SDR}
    S(\omega)-S(-\omega)=2i\lambda \sinh\frac{\omega}{2T},
\end{equation}
where $S(\omega)$ is defined as an infinite product
\begin{equation}
    S(\omega)=\qty(1+\frac{\omega}{\omega_*})\prod_n \qty(1-\frac{\omega}{\omega_n^+})\qty(1+\frac{\omega}{\omega_n^-}), \label{def:sasian}
\end{equation}
which, again, should be taken with respect to concentric disks centred at the origin, as described above. The function $S(\omega)$ contains all the information about the longitudinal- and transverse-channel spectra. Specifically, its zeroes in the lower complex half-plane describe the longitudinal spectrum, whereas its zeroes in the upper complex half-plane describe the transverse spectrum. The constant-in-$\omega$ function $\lambda(k)$ in Eq.~\eqref{eq:SDR} is related to the constant-in-$\omega$ functions appearing in the thermal product formula $\eqref{def:TPF}$ via
\begin{equation}
    \lambda=\frac{\mu_+}{\alpha}=\alpha \mu_-, \label{eq:lambdaEq}
\end{equation}
or, stated differently.
\begin{align}
    \lambda^2=\mu_+\mu_-.
\end{align}
While $\mu$ and $\alpha$, both being Taylor coefficients of retarded correlators expanded around $\omega=0$, can be rescaled, $\lambda$ is completely defined by the spectra. In fact, in Section~\ref{subsec:neutral}, we show how $\lambda$ can be expressed through the spectrum of only a single channel.

The spectral duality relation \eqref{eq:SDR} presents an incredibly stringent, infinite set of constrains (i.e., `one' for each $\omega$) between the spectra of the longitudinal and transverse channels. It is immediately clear that one can express \emph{any} finite number of QNMs, if the remaining (infinitely many) QNMs are known. Furthermore, the spectral duality relation is robust in the sense of being independent of the detailed definitions of $G_\pm(\omega)$. The main input from the bulk is the algebraically special frequency $\omega_*$, which appears in the definition of $S(\omega)$ in Eq.~\eqref{def:sasian}.

Remarkably, starting from the spectrum of one correlator, say, the longitudinal correlator, the above construction allows us to compute the residues from Eq.~\eqref{eq:residues}. This fixes the retarded correlator. Then, we can use the duality relation \eqref{eq:DUALITY} to compute the spectrum of the second correlator, i.e., the transverse correlator. We show how this can be done  in detail for the case of the energy-momentum correlators in a thermal (neutral) state in Section~\ref{subsubsec:energy-momentum}. For further, more general discussions of related matters, we refer the reader to Ref.~\cite{GV:2025}.

\subsection{The Schwarzschild black brane and the neutral thermal CFT state}
\label{subsec:neutral}
We now begin with our study of the 3$d$ thermal boundary CFT state, dual to the 4$d$ asymptotically-AdS Schwarzschild black brane of Section~\ref{subsubsec:schw} (the case with $K=0$). The only scale in the problem is the temperature $T$. In this setup, we can express the boundary expectation values of conserved operators in terms of the bulk curvature tensors (see Refs.~\cite{Mansi:2008br,Compere:2008us,Bakas:2009da}). More specifically, it is convenient to write the boundary U$(1)$ current and the energy-momentum tensor with the electric components of the curvature tensors (see Eqs.~\eqref{def:Emaxwell} and \eqref{def:Eweyl})
\begin{subequations}
\label{def:currents}
\begin{align}
    \expval{J^{a}} &=-\gamma_J \lim_{r\rightarrow\infty}r^3 E^a(r), \label{def:jU}\\
    \expval{T^{ab}}&=-\gamma_T \lim_{r\rightarrow\infty} r^5 E^{ab}(r), \label{def:Tmunu}
\end{align} 
\end{subequations}
where $\gamma_J$ and $\gamma_T$ are some (arbitrary) normalisation constants. We also note that, provided that the Einstein's equations are solved, there is no need to renormalise the Weyl tensor with any counterterms to obtain the boundary energy-momentum tensor, and the definition of Eq.~\eqref{def:Tmunu} is, at least in the linearised sense, equivalent to the standard definition in terms of the Brown-York tensor \cite{Balasubramanian:1999re,Skenderis:2000in}. For the Schwarzschild case, the boundary energy density is expressed as
\begin{equation}
    \overline{\epsilon} = \expval{T^{tt}}_0 =2M \gamma_T  = \gamma_T\qty(\frac{4 \pi T}{3})^3,
\end{equation}
where $\expval{...}_0$ represents the expectation value in the absence of external sources. The boundary charge density is zero in this (neutral) state, i.e., 
\begin{equation}
    \overline{n} = \expval{J^t}_0 = 0.
\end{equation}
The following (linearised) Ward identities hold:
\begin{subequations}
\label{eq:Ward}
    \begin{align}
    \nabla_a \expval{T^{ab}}&=0,\\
    g^{(0)}_{ab}\expval{T^{ab}}&=0,\\
    \nabla_a \expval{J^{a}}&=0, \label{eq:JJ_Ward}
\end{align}
\end{subequations}
where $\nabla_a$ is the covariant derivative compatible with $g_{ab}^{(0)}$. Here, we have used the fact that, in the neutral state, the current and the energy-momentum tensor decouple. Taking variational derivatives of Eqs.~\eqref{eq:Ward} then gives the Ward identities for the retarded two-point correlators (see e.g.~Refs.~\cite{Policastro:2002tn,Herzog:2009xv}). Since we are considering an isotropic CFT state, both the current and the energy-momentum correlators are completely determined by their respective longitudinal- and transverse-channel correlators \cite{Kovtun:2005ev}.

Instead of working with the boundary metric and the gauge field, it will prove convenient to work with the respective `topological currents' associated with the external fields. For the external gauge field, this is the (three-dimensional) Hodge dual of the associated boundary field strength tensor, and for the metric, it is the Cotton-York tensor $C^{ab}$:
\begin{subequations}
\label{def:sources}
\begin{align}
    C^a&=\epsilon^{abc}\partial_b A_c^{(0)}, \label{def:topological}\\
    C^{ab}&=\nabla_k \qty(R_{l}{}^{a}[g^{(0)}]-\frac{1}{4} \delta_{l}{}^{a}R[g^{(0)}] ) \epsilon^{kl}{}_b. \label{def:cotton}
\end{align} 
\end{subequations}
$R^{ab}[g^{(0)}]$ and $R[g^{(0)}]$ are the boundary Ricci tensor and scalar, respectively. Both $C^a$ and $C^{ab}$ are invariant with respect to the gauge transformations and are trivially conserved for all configurations. Furthermore, the Cotton-York tensor $C^{ab}$ is traceless and invariant under the conformal transformations of the boundary metric. Consequently, $C^a$ and $C^{ab}$ encode all the gauge-invariant and conformally-invariant information about the linearised boundary sources.\footnote{They do not, however, contain the information about the pure gauge terms, such as the boundary chemical potential.} The main convenience of using $C^a$ and $C^{ab}$ over $A^{(0)}_a$ and $h^{(0)}_{ab}$ lies is the fact that they can be directly expressed via the magnetic parts of the bulk field curvature tensors (see Eqs.~\eqref{def:Bmaxwell} and \eqref{def:Bweyl}). Namely,
\begin{subequations}
\label{def:sourced_holo}
\begin{align}
    C^a&=\lim_{r\rightarrow\infty}r^3 B^a(r),\\
    C^{ab}&=\lim_{r\rightarrow\infty} r^5 B^{ab}(r). \label{def:bdry_cotton}
\end{align}    
\end{subequations}
The fact that the topological current and the Cotton-York tensor exhibit the same properties as the conserved U$(1)$ current and the energy-momentum tensor lies at the heart of why 3$d$ CFTs enjoy a breadth of dualities.

\subsubsection{The current correlators}
\label{subsubsec:JJ}
Next, we consider the retarded two-point correlators of the conserved U$(1)$ current. We will use notation that will make the transition to the discussion of the energy-momentum tensor natural. We have (we henceforth omit the expectation value notation for readability)
\begin{subequations}
\begin{align}
    \CE^{(0)}_+ &= -\frac{1}{\gamma_J}\delta J^z,\\
    \CE^{(0)}_- &= -\frac{1}{\gamma_J}\delta J^y,\\
    \CB^{(0)}_+ &= C^y=-ik  A^{(0)}_t-i\omega  A^{(0)}_z,\\
    \CB^{(0)}_- &= C^z=i\omega  A^{(0)}_y,
\end{align}    
\end{subequations}
where $\CE_\pm^{(0)}$ and $\CB_\pm^{(0)}$ are the boundary values of $ik\CE_\pm(r)$ and $ik\CB_\pm(r)$:
\begin{align}
    \CE^{(0)}_\pm&=ik\lim_{r\rightarrow\infty} \CE_\pm(r), \\
    \CB^{(0)}_\pm&=ik\lim_{r\rightarrow\infty} \CB_\pm(r).
\end{align}
We define the longitudinal and the transverse channel current-current correlators as
\begin{subequations}
\label{def:JJ_correlators}
\begin{align}
    G^J_+ &= -\frac{1}{i\omega}\frac{\CE^{(0)}_+}{\CB^{(0)}_+},\\
    G^J_- &= -i\omega \frac{\CE^{(0)}_-}{\CB^{(0)}_-}.
\end{align}    
\end{subequations}
Note that, since they are gauge invariant, $\CE_\pm$ must be linear in $\CB_\pm$ and thus the correlators $G^J_\pm$ are well defined as simple fractions. The factors of $i\omega$ assure that $G^J_\pm$ have no poles or zeroes at $\omega=0$ for any given finite $k$, which can be inferred from the Ward identities, as well as from the positivity \eqref{eq:positivity}. Given Eqs.~\eqref{eq:dualityEM}, the correlators $G^J_\pm$ obey the self-duality relation as given in Eq.~\eqref{eq:SELF_DUALITY}. One can use the Ward identity \eqref{eq:JJ_Ward} to express $G^J_\pm$ in the usual way as
\begin{subequations}
\begin{align}
    G^J_+ &= \frac{1}{\gamma_J k^2}\frac{\delta \expval{J^t}}{\delta A_t}, \\
    G^J_- &= \frac{1}{\gamma_J}\frac{\delta \expval{J^y}}{\delta A_y}.
\end{align}    
\end{subequations}

Before proceeding, we note that the self-duality relation of the current correlators is in QFT terms related to two distinct features of the AdS$_4$/CFT$_3$. Firstly, the swapping of the Dirichlet (magnetic) boundary conditions with the Neumann (electric) boundary conditions corresponds to the particle-vortex duality (or the S-duality) of the boundary theory \cite{Witten:2003ya}. Secondly, the self-duality of the bulk Maxwell equations tells us that the retarded two-point correlators of the particle-vortex dual theories are the same \cite{Herzog:2007ij}. 

\begin{figure}
    \centering
    \includegraphics[width=\linewidth]{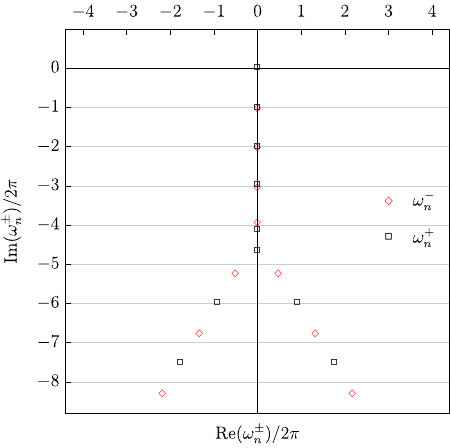}
    \caption{The poles of $G_+$ (black squares) and $G_-$ (red diamonds) at $k=0.1$. In both channels, the poles converge to the Matsubara frequencies at zero momentum. The longitudinal channel exhibits a hydrodynamic charge diffusion mode.}
    \label{fig:JJ_evenodd}
\end{figure}

\begin{figure}
    \centering
    \includegraphics[width=\linewidth]{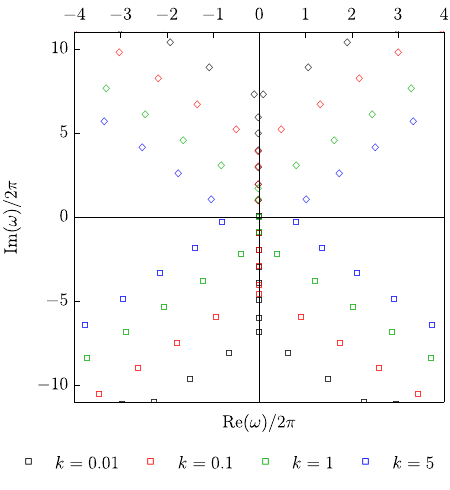}
    \caption{The zeroes of $S(\omega)$ associated with $G_\pm^J$. As we lower momentum, the asymptotic branches collide with one another, and poles converge to the Matsubara frequencies in the zero-momentum limit.}
    \label{fig:JJ_S}
\end{figure}

We now turn to the spectra of $G_\pm^J$. At large momenta, the spectra in both channels are organised into two asymptotic `Christmas tree' lines. As we lower $k$, the poles converge to the imaginary axis in a `zipper-like' manner in through a sequence of pair-wise pole collisions, as shown in Figure~\ref{fig:JJ_evenodd} (see also Ref.~\cite{Witczak-Krempa:2012qgh}). The longitudinal channel features a hydrodynamic charge diffusion mode $\omega_\text{diff}$, with the dispersion relation
\begin{equation}
    \omega_\text{diff}(k^2)=-i D_c k^2 +\CO(k^4).
\end{equation}
The rest of the modes are gapped, and, at finite momentum, organise themselves into two mirrored asymptotic branches. The function $S(\omega)$ contains all the information about poles in both channels, as is shown in Figure~\ref{fig:JJ_S}. Note that the diffusive mode acquires a real part upon collision with an imaginary pole.
We first focus on the zero-momentum limit. In this case, the correlators can be computed explicitly \cite{Herzog:2007ij}. This is due to the self-duality relation \eqref{eq:SELF_DUALITY} and isotropy, which gives
\begin{equation}
    \left.\frac{G_-^J}{G_+^J}\right|_{k=0}=\omega^2.
\end{equation}
Hence,
\begin{subequations}
\label{eq:JJ_k0}
\begin{align}
    G_+^J(\omega,k=0)=-\frac{1}{i\omega},\\
    G_-^J(\omega,k=0)=i\omega.
\end{align}
\end{subequations}
As already noted in Ref.~\cite{Grozdanov:2024wgo}, another way to see this is by using the spectral duality relation. We can write
\begin{equation}
    S(\omega,k)=\qty(1-\frac{\omega}{\omega_\text{diff}})S_\text{gap}(\omega,k),
\end{equation}
where $S_\text{gap}$ contains the information about all gapped (non-hydrodynamic) modes. Due to isotropy at $k=0$,
\begin{equation}
    S_\text{gap}(\omega,0)=S_\text{gap}(-\omega,0).
\end{equation}
Expanding the spectral duality relation for small $k$, we obtain the limit 
\begin{equation}
    \frac{i D_c}{2T}\lim_{k\rightarrow 0}k^2 S(\omega,k) =\sinh \frac{\omega}{2T},
\end{equation}
which enforces that the gapped modes in both channels limit to the multiples of the negative imaginary Matsubara frequencies as $k\to 0$. In particular,  
\begin{subequations}
    \begin{align}
    \omega_n^+(k=0) &= -2\pi i n T, \qquad n=0,1,2,\ldots,\\
    \omega_n^-(k=0) &= -2 \pi i n T, \qquad n=1,2,3,\ldots.
\end{align}
\end{subequations}
This makes it transparent that the explicit expressions for the correlators at zero momentum \eqref{eq:JJ_k0} are just the product expansions from Eq.~\eqref{eq:Gproducts} with all the gapped poles cancelling each other. Indeed, this is the only way that the expressions in Eq.~\eqref{eq:JJ_k0} can be compatible with the thermal product formula, given that the only way for the residues to vanish is for the poles to be precisely at the Matsubara frequencies. This is related to the phenomenon of pole skipping. This behaviour of QNMs is universal in the self-dual limit of isotropic systems (at $k=0$) with one gapless mode, and can be seen, for example, in the self-dual limit of the linear axion model as well. We can also deduce the low-$k$ behaviour of the parameter $\lambda$ purely from the diffusive mode, i.e.,
\begin{equation}
    \lambda =-\frac{2T}{D_c k^2}+\CO(k^0).
\end{equation}

We now turn to finite $k$. Suppose we work at a large enough $k$ so that there are no purely damped (imaginary) QNMs in the spectrum. For convenience, we denote the set of all the poles with $\Re(\omega_n)>0$ by $\widetilde\omega_n^\pm$, in order of ascending modulus. At a given momentum, the poles are organised into asymptotic lines (see Figure~\ref{fig:JJ_S})
\begin{subequations}
\begin{align}
    \omega_n^\text{asymp}=r n + s,
\end{align}
\end{subequations}
so that
\begin{subequations}
\begin{align}
    \widetilde\omega_n^\pm &= \omega_n^\text{asymp} +\delta^\pm_\text{asymp}(n),
\end{align}
\end{subequations}
where $r$ and $s$ are some complex parameters, and $\delta^\pm_\text{asymp}(n)<C$ for some constant $C$. It can be checked (see Figure~\ref{fig:JJ_evenodd}) that there is a relative asymptotic displacement between the longitudinal and transverse channels that behaves as
\begin{equation}
    \lim_{n\rightarrow \infty}\qty[\delta^+_\text{asymp}(n)-\delta^-_\text{asymp} (n)] = -\frac{r}{2}.
\end{equation}
This can then be used to understand the large-$\omega$ asymptotics of the product expansions \eqref{eq:Gproducts}. In particular (see Ref.~\cite{GV:2025} for details),
\begin{subequations}
    \begin{align}
        \omega &\rightarrow \infty:& G^J_+(\omega) &\sim \frac{1}{\omega},\\
        & & G^J_-(\omega) &\sim \omega.
    \end{align}
\end{subequations}
It is straightforward to see that the same asymptotic behaviour is true for all momenta. This allows for the following form of a partial fraction decomposition (see e.g.~Ref.~\cite{smirnov}) of both meromorphic correlators (see also Ref.~\cite{Casalderrey-Solana:2018rle}):
\begin{subequations}
    \begin{align}
    G^J_+(\omega)&=2i\sum_n \frac{r_n^+}{\omega-\omega_n^+},\\
    G^J_-(\omega)&=g_0^- + \mu_-\frac{i\omega}{2T}+2i\omega^2 \sum_n \frac{r_n^-}{(\omega_n^-)^2}\frac{1}{\omega-\omega_n^-},
\end{align}
\end{subequations}
where the residues are expressed via Eq.~\eqref{eq:residues} as
\begin{equation}
    r^\pm_n = -\mu_\pm \frac{\omega^\pm_n \sinh\frac{\omega^\pm_n}{2T}}{2 \prod\limits_{m \neq n}\qty(1-\qty(\frac{\omega^\pm_n}{\omega^\pm_m})^2)}.
\end{equation}
$G^J_+$ is completely determined, up to a $k$-dependent scaling constant $\mu_+$. This allows us to express $\lambda(k)$ in terms of the longitudinal spectrum
\begin{equation}
    \frac{1}{\lambda(k)}=-i \sum_n \frac{\sinh\frac{\omega^+_n}{2T}}{\prod\limits_{m\neq n}\qty[1-\qty(\frac{\omega^+_n}{\omega^+_m})^2]},
\end{equation}
and also find a specific constraint on the longitudinal spectrum by matching the small-$\omega$ expansions of both sides of the thermal product formula \eqref{def:TPF},
\begin{equation}
     \sum_n \frac{\sinh\frac{\omega^+_n}{2T}}{\omega_n^+\prod\limits_{m\neq n}\qty[1-\qty(\frac{\omega^+_n}{\omega^+_m})^2]} = \frac{1}{2T}.
\end{equation}
This expression holds for any $k$ and is equivalent to the sum rule derived in our Ref.~\cite{Grozdanov:2024wgo}. With the knowledge of the longitudinal spectrum, we can now derive the transverse spectrum by solving
\begin{equation}
    G^J_+(\omega_n^-)=0.
\end{equation}
Conversely, we can derive the longitudinal spectrum from the transverse spectrum by solving
\begin{equation}
    G^J_-(\omega_n^+)=0,
\end{equation}
where we have to first fix the constant $g_0^-$. This can be done, for example, with the knowledge of one mode from the longitudinal channel. These properties then allow us to determine each of the spectra from the other, dual spectrum. We do not present the numerical details of this construction here, but instead focus on the more interesting case of the energy-momentum correlators in Section~\ref{subsubsec:numerics}.

\subsubsection{The energy-momentum tensor correlators: definitions}
\label{subsubsec:energy-momentum}
The case of the energy-momentum tensor dualities is both more interesting and more subtle due to the operators' finite equilibrium expectation value. Here, we can express the relevant boundary variables as 
\begin{subequations}
\begin{align}
    E^{(0)}_+&=\frac{1}{4\gamma_T}\left[  
 2(\delta T_{zz}-\delta T_{yy}) 
 \right. \nonumber\\
 &\left.+\overline{\epsilon}(\delta g_{yy}-\delta g_{zz}) \right] , \\
    E^{(0)}_-&=\frac{1}{2\gamma_T}\qty(2 \delta T_{yz}-\overline{\epsilon}\delta g_{yz}),\\
    B^{(0)}_+&=\frac{i\omega}{4}\left[k^2 \delta g_{tt}+2\omega k \delta g_{tz} \right. \nonumber\\
    &+ \left. \omega^2\delta g_{zz}+(k^2-\omega^2)\delta g_{yy}\right] ,\\
    B^{(0)}_-&=\frac{k^2-2\omega^2}{4}(i k \delta g_{ty}+i\omega\delta g_{yz}),
\end{align}
\end{subequations}
where $E^{(0)}_\pm$ and $B^{(0)}_\pm$ are the boundary values of $E_\pm$ and $B_\pm$,
\begin{subequations}
\begin{align}
    E^{(0)}_\pm &= \frac{1}{2}k^2\lim_{r\rightarrow\infty}E_\pm(r), \\
    B^{(0)}_\pm &= \frac{1}{2}k^2\lim_{r\rightarrow\infty}B_\pm(r).
\end{align}    
\end{subequations}
The prefactors are again set by the normalisation of the harmonics.

We can now define the longitudinal and transverse channel energy-momentum tensor correlators as
\begin{subequations}
\label{def:naturalGpm}
 \begin{align}
    G^T_+&=\frac{2i\omega}{k^2-2\omega^2}\qty(\frac{ E^{(0)}_+}{ B^{(0)}_+}+\frac{\omega}{\omega_*}),\\
    G^T_-&=\frac{k^2-2\omega^2}{2i\omega}\qty(\frac{ E^{(0)}_-}{ B^{(0)}_-}+\frac{\omega}{\omega_*}),
\end{align}   
\end{subequations}
where the prefactors are included to avoid spurious zeroes or poles, guaranteeing that $G^T_\pm$ have no poles or zeroes at $\omega=0$ or $\omega^2=k^2/2$ for non-zero momenta. Using the duality relations from Eqs.~\eqref{eq:dualityEMmetric}, we can conclude that $G^T_+$ and $G^T_-$ must obey the duality relation \eqref{eq:DUALITY}, with $\omega_*$ defined as
\begin{equation}
    \omega_*=i \frac{\gamma_T k^4}{6\overline{\epsilon}},
\end{equation}
which is equivalent to the expression from Eq.~\eqref{def:AS}. Note that the definitions \eqref{def:naturalGpm}, along with the algebraically special conditions \eqref{eq:EBalgSpec}, immediately imply that $G^T_+(\omega_*)=G^T_-(-\omega_*)=0$. Next, by taking the Ward identities \eqref{eq:Ward} into account, we can express $G^T_+$ and $G^T_-$ in a more familiar but less transparent way: 
\begin{align}
    G^T_+&=\frac{4}{\gamma_T k^4}\qty[\frac{\delta \sqrt{-g} \expval{T^{tt}}}{\delta g_{tt}}-2\bar\epsilon],\label{def:TT_plus} \\
    G^T_-&=\frac{4}{\gamma_T k^4}\qty[k^2\frac{\delta \sqrt{-g} \expval{T^{ty}}}{\delta g_{ty}}+\bar\epsilon\frac{6 \omega^2-k^2}{4}]\label{def:TT_minus}.
\end{align}

The explicit expressions in Eqs.~\eqref{def:naturalGpm} bring us a step closer to understanding what dualities in the form of Eq.~\eqref{eq:DUALITY} mean in thermal field theory. In the `self-dual' pure AdS limit, $M\rightarrow 0 $, we can use the Lorentz invariance to conclude that 
\begin{equation}
\left.\frac{G^T_-(\omega,k)}{G^T_+(\omega,k)}\right|_{\overline\epsilon =0} = \frac{\omega^2-k^2}{2},
\end{equation}
which, in conjunction with the self-duality relation \eqref{eq:SELF_DUALITY}, gives the conformally-invariant forms of the correlators
\begin{subequations}
    \begin{align}
        \left(G^T_+\right)^2&=\frac{k^2-\omega^2}{2},\\
        \left(G^T_-\right)^2&=\frac{2}{k^2-\omega^2}.
    \end{align}
\end{subequations}
This self-duality is the gravitational analogue of the S-duality, where one lets the boundary metric vary, which exchanges the conserved energy-momentum tensor with the conserved Cotton-York tensor \cite{Leigh:2003ez,Bakas:2008gz,Bakas:2009da,deHaro:2008gp,Compere:2008us}. Holographically, this again amounts to switching between the Dirichlet and Neumann boundary conditions for the metric. Turning on the temperature is accompanied by the emergence of the algebraically special structure, which, at present, is not understood in terms of the boundary QFT. We note that the expressions \eqref{def:naturalGpm}, in conjunction with the duality relation \eqref{eq:DUALITY}, can be interpreted as the fact that, at the level of the linearised theory, the Dirichlet boundary conditions ($B^{(0)}_\pm=0$) are equivalent to mixed boundary conditions:
\begin{subequations}
 \begin{align}
    B^{(0)}_+ &= 0 ~~ \Leftrightarrow ~~ \frac{\omega B^{(0)}_-+\omega_* E^{(0)}_-}{\omega+\omega_*}=0, \\
    B^{(0)}_- &= 0 ~~ \Leftrightarrow ~~ \frac{\omega B^{(0)}_++\omega_* E^{(0)}_+}{\omega-\omega_*}=0.
\end{align}   
\end{subequations}

\subsubsection{The energy-momentum tensor correlators: spectra}
\label{subsubsec:numerics}

In this section, we discuss the spectra of the  energy-momentum correlators in the neutral thermal state in detail, and present various numerical results.\footnote{We have used the \texttt{QNMSpectral} Mathematica library \cite{Jansen:2017oag} to aid with the numerical computations.} We set the temperature to $T=1$ throughout. 

\begin{figure*}[th!]
    \centering
    \includegraphics[width=0.49\textwidth]{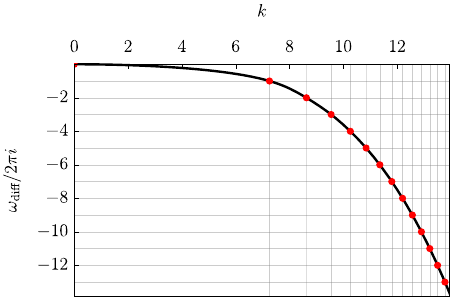}
    \hfill
    \includegraphics[width=0.49\textwidth]{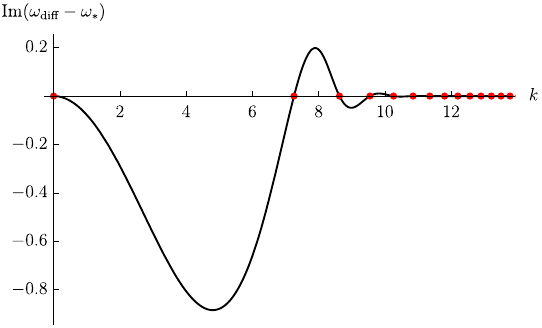}
    \caption{Left panel: The diffusive mode interpolates the pole-skipping points from Eq.~\eqref{eq:PS2} at integer multiples of the negative imaginary Matsubara frequencies. Right panel: The difference $\omega_\text{diff}-\omega_*$ vanishes at the pole-skipping points, and rapidly decays to zero as real $k$ is increased. This implies that the function $-\omega_*(k)$ can be used as an excellent large-$k$ approximation of the diffusive dispersion relation.}
     \label{fig:pole-skipping}
\end{figure*}

The spectra of both channels split into their respective hydrodynamic and gapped parts. In the longitudinal channel, there are no purely imaginary QNMs, therefore all of them come in mirrored pairs of $(\omega_n, -\omega_n^*)$. The spectrum contains the hydrodynamic sound mode (see e.g.~Ref.~\cite{Grozdanov:2019kge})
\begin{equation}
    \omega_\text{sound} = v_s k-i\Gamma k^2 +\CO(k^3), \label{eq:sound_mode}
\end{equation}
and an infinite number of gapped modes $\omega_{\text{gap},n}^+$, each with its respective mirrored mode $(-\omega_{\text{gap},n}^+)^*$. In Eq.~\eqref{eq:sound_mode}, $v_s$ is the speed of sound and $\Gamma$ the attenuation rate. In the transverse channel, we have one hydrodynamic mode---the imaginary diffusive mode with the diffusivity $D$, 
\begin{equation}
    \omega_\text{diff} = -iDk^2+\CO(k^4), \label{eq:diff_mode}
\end{equation}
and an infinite number of gapped QNMs $\omega_{\text{gap},n}^-$ with their respective mirrored modes $(-\omega_{\text{gap},n}^-)^*$. As real $k$ increases, the diffusive mode interpolates through the infinite sequence of pole-skipping points at (see Refs.~\cite{Grozdanov:2020koi,Grozdanov:2023txs})\footnote{See also Ref.~\cite{Grozdanov:2023tag} for questions related to this interpolation problem and the reconstruction of the diffusive mode from the pole-skipping points.}
\begin{equation}
    \omega=-2\pi i n T, \quad k = \frac{4 \pi T}{\sqrt{3}} n^{1/4}, \label{eq:PS2}
\end{equation}
for $n=0,1,2,\ldots$, as shown in Figure~\ref{fig:pole-skipping}.

\begin{figure}[ht!]
    \centering
    \includegraphics[width=\linewidth]{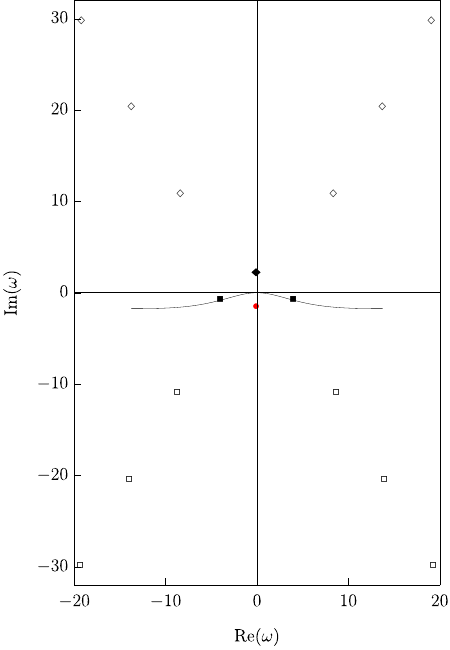}
    \caption{The zeroes of $S(\omega)$ for the energy-momentum tensor correlators at $k=5$. The lower complex half-plane contains the longitudinal spectrum (squares), whereas the upper complex half-plane contains the transverse spectrum (diamonds). The set of all zeroes is split into the hydrodynamic modes (filled), gapped modes (empty), and the algebraically special frequency (red dot), which acts as a `faux longitudinal mode'. The dependence of the sound modes on real $k \in [0,14]$ is shown with a black line.}
    \label{fig:TT_S}
\end{figure}

\begin{figure}
    \centering
    \includegraphics[width=\linewidth]{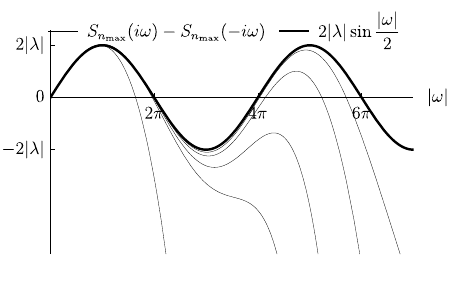}
    \caption{Numerical verification of the spectral duality relation \eqref{eq:SDR} for $k=5$ and $n_\text{max}=0,3,5,10,20$. Here, $\omega$ is imaginary for illustrative purposes. It is clear that the odd part of $S_{n_\text{max}}(\omega)$ converges to the sine function, satisfying the spectral duality relation.}
    \label{fig:S_convergence}
\end{figure}

The complete information about both channels' QNM spectra is again contained in $S(\omega)$, as defined in Eq.~\eqref{def:sasian}. The zeros of the function are shown in Figure~\ref{fig:TT_S}.
The spectral duality relation can be explicitly verified by using the numerically computed QNMs. To show this, suppose we have the knowledge of all the hydrodynamic modes and $n_\text{max}$ lowest-laying gapped modes in both channels. Here, $n_\text{max}$ counts the modes only in one quadrant of the complex plane, so the total number of the modes is $3+4 n_\text{max}$. We can then approximate $S(\omega)$ by 
\begin{equation}
    S_{n_\text{max}}(\omega)=\qty(1+\frac{\omega}{\omega_*})S^\text{hy}(\omega)S^\text{gap}_{n_\text{max}}(\omega),
\end{equation}
where 
\begin{subequations}
    \begin{align}
        \hspace{-0.5cm}S^\text{hy}(\omega)&=
        \qty(1-\frac{\omega}{\omega_\text{sound}})
        \qty(1+\frac{\omega}{\omega^*_\text{sound}})
        \qty(1+\frac{\omega}{\omega_\text{diff}}), \hspace{-0.5cm}  \\
        \hspace{-0.5cm}S^\text{gap}_{n_\text{max}}(\omega)&=\prod_{n=1}^{n_\text{max}}
        \qty(1-\frac{\omega}{\omega_{\text{gap},n}^+})
        \qty(1+\frac{\omega}{{(\omega_{\text{gap},n}^+)}^*}) \nonumber\hspace{-0.5cm} \\
    \hspace{-0.5cm}&\times\qty(1+\frac{\omega}{\omega_{\text{gap},n}^-})
        \qty(1-\frac{\omega}{{(\omega_{\text{gap},n}^-)}^*}).\hspace{-0.5cm}
    \end{align}
\end{subequations}
In the limit of $n_\text{max}\rightarrow \infty$, the odd part of $S_{n_\text{max}}(\omega)$ obeys the spectral duality relation \eqref{eq:SDR}, as demonstrated in Figure~\ref{fig:S_convergence}. 

With the hydrodynamic dispersion relations from Eqs.~\eqref{eq:sound_mode} and \eqref{eq:diff_mode} in hand, one can now expand the spectral duality relation for small momenta and get a fixed ratio between the diffusivity and the attenuation rate of the sound mode \cite{Grozdanov:2024wgo}:
\begin{equation}
    \frac{D}{\Gamma}=2.
\end{equation}
This can also be seen as a statement about the vanishing bulk viscosity. While this is the case in all 3$d$ CFTs that admit a hydrodynamic description, the result here does not rely on a specific construction of the hydrodynamic low-energy effective theory. One can similarly derive the low-energy behaviour of the parameter $\lambda$ as
\begin{equation}
    \lambda = \frac{2 T}{i\omega_*} + \CO(k^{-2}).
\end{equation}

The asymptotic behaviour of the gapped modes is the same in both channels, with
\begin{align}
    \omega_{\text{gap},n}^\pm = \omega_n^\text{asymp}+\delta^\pm_\text{asymp}(n),
\end{align}
where $\omega_n^\text{asymp}$ is known analytically from calculations using the WKB approximation \cite{Cardoso:2004up,Natario:2004jd}. To subleading order in large $n$, 
\begin{equation}
\label{eq:WKB}
    \omega_n^\text{asymp}=\pi T \qty[(\sqrt{3}-3i) n+1-i\frac{\ln4}{\pi}],
\end{equation}
and it can be checked that $\delta^\pm_\text{asymp}(n)$ vanishes for $n\rightarrow\infty$. Then, using the product expansion of retarded correlators \eqref{eq:Gproducts}, it follows that at large (complex) $\omega$, we have
\begin{subequations}
    \begin{align}
        \omega &\rightarrow \infty:& G^T_+(\omega) &\sim \omega^0,\\
        & & G^T_-(\omega) &\sim \omega^2,
    \end{align}
\end{subequations}
which can also be understood from the low-energy part of the spectrum, since the asymptotic gapped modes effectively cancel one another when considering large $\omega$ (see Ref.~\cite{GV:2025} for details). Therefore we can express the retarded correlators in terms of the partial fraction decomposition:
\begin{subequations}
\label{eq:partial_fractions}
    \begin{align}
    G^T_+(\omega)&=2i   \sum_n \frac{r_n^+}{\omega_n}\qty[\frac{\omega}{\omega-\omega_n^+}-\frac{\omega_*}{\omega_*-\omega_n^+}],\label{eq:Geven_partial}\\
    G^T_-(\omega)&= \frac{i\mu_-}{2T}(\omega+\omega_*)+g^-_2 (\omega^2-\omega_*^2)+\\&+2i  \sum_n \frac{r_n^-}{(\omega_n^-)^3}\qty[\frac{\omega^3}{\omega-\omega_n^-}-\frac{\omega_*^3}{\omega_*+\omega_n^-}],
\end{align}
\end{subequations}
with the residues again expressed via the thermal product formula (cf.~Eq.~\eqref{eq:residues}):
\begin{equation}
    r^\pm_n = -\mu_\pm \frac{\omega^\pm_n \sinh\frac{\omega^\pm_n}{2T}}{2 \prod\limits_{m \neq n}\qty(1-\qty(\frac{\omega^\pm_n}{\omega^\pm_m})^2)}.
\end{equation}

The longitudinal channel correlator is completely determined by one $k$-dependent parameter $\mu_+$ that appears in the thermal product formula. On the other hand, the transverse channel correlator contains a $k$-dependent real parameter $g_2^-$, which has to be fixed by some independent means and therefore requires `more information' to specify than the longitudinal correlator. Focusing on the longitudinal correlator, one can expand Eq.~\eqref{eq:Geven_partial} for small $\omega$ and use Eq.~\eqref{eq:lambdaEq} to express $\lambda$ purely in terms of $\omega_*$ and the longitudinal spectrum,
\begin{equation}
    \frac{1}{\lambda(k)}=-i \sum_n \frac{\sinh\frac{\omega^+_n}{2T}}{\qty(1-\frac{\omega^+_n}{\omega_*})\prod\limits_{m\neq n}\qty[1-\qty(\frac{\omega^+_n}{\omega^+_m})^2]}.
\end{equation}
One can similarly write a (sum rule \cite{Grozdanov:2024wgo}) constraint on the longitudinal spectrum that holds for all $k$ as
\begin{equation}
     \sum_n \frac{\sinh\frac{\omega^+_n}{2T}}{\omega_n^+\prod\limits_{m\neq n}\qty[1-\qty(\frac{\omega^+_n}{\omega^+_m})^2]} = \frac{1}{2T}.
\end{equation}

We now demonstrate how to express the spectrum in one channel in terms of the spectrum from the other channel. To do so, we use the partial fraction decompositions of the correlators (cf.~Eqs.~\eqref{eq:partial_fractions}). Suppose we have the knowledge of the hydrodynamic modes as well as the first $n_\text{max}$ gapped modes. As was noted in Ref.~\cite{Dodelson:2023vrw}, the convergence of the calculation in terms of $n_\text{max}$ radically improves by taking into account the asymptotic expansion in Eq.~\eqref{eq:WKB}. While there are several ways of approaching the problem, we present what we have found to be particularly convenient.

We first focus on the spectral function $\rho(\omega)$ (this can be done in either of the channels). We approximate the unknown modes by using the asymptotic expansion \eqref{eq:WKB}, i.e.,
\begin{equation}
    \omega_{\text{gap},n}\approx \omega^\text{asymp}_n, \qquad n>n_\text{max}. \label{eq:w_approx}
\end{equation}
Now let $\CA$ denote the set of the first $2q+n_\text{hy}$ lowest-laying modes, where $n_\text{hy}$ is the number of the hydrodynamic modes. That means that it includes $\omega_{\text{gap},q}$ and its mirror image, but no modes with a larger modulus. We can then approximate the spectral function in a way which takes into account the infinite number of asymptotic modes. Specifically, we can express the `truncated' spectral function as 
\begin{equation}
    \rho^T (\omega)\approx\frac{H(\omega;n_\text{max})}{H(0;n_\text{max})}\frac{\mu \sinh\frac{\omega}{2T}}{\prod\limits_{\omega_n \in \CA}\qty(1-\frac{\omega^2}{\omega_n^2})},
\end{equation}
where
\begin{align}
H(\omega;n_\text{max})&=\Gamma\qty(\frac{A+\omega}{B}+1)\Gamma\qty(\frac{A-\omega}{B}+1) \nonumber \\ &\times\Gamma\qty(\frac{A^*+\omega}{B^*}+1)\Gamma\qty(\frac{A^*-\omega}{B^*}+1),
\end{align}
with
\begin{subequations}
\begin{align}
    A &= \omega^\text{asymp}_{n_\text{max}}, \\
    B &= \pi T(\sqrt{3}-3i).
\end{align}
\end{subequations}
The residues of the known modes are then well approximated with
\begin{equation}
    r_n \approx -\mu \frac{H(\omega_n;n_\text{max})}{H(0;n_\text{max})}\frac{\omega_n \sinh\frac{\omega_n}{2T}}{2 \prod\limits_{\omega_m \in \CA \backslash \qty{\omega_n}}\qty(1-\frac{\omega_n^2}{\omega_m^2})}. \label{eq:rn_approx}
\end{equation}
Next, we use the approximate modes \eqref{eq:w_approx} with their respective residues \eqref{eq:rn_approx} in the expansions \eqref{eq:partial_fractions}. We can then express the transverse spectrum $\omega_n^-$ by numerically solving the following equation:
\begin{equation}
    \frac{G^T_+(\omega_n^-)}{\omega-\omega^*}=0. \label{eq:odd_compute}
\end{equation}
Note that this expression is independent of $\mu_+$, therefore the knowledge of the longitudinal spectrum is enough to completely determine the transverse spectrum. The transverse correlator, on the other hand, has a free constant $g_-(k)$, which, as discussed above, has to be fixed by independent means. The longitudinal spectrum $\omega_n^+$ is then determined by the transverse spectrum by numerically solving
\begin{equation}
    \frac{G^T_-(\omega_n^+)}{\omega+\omega^*}=0.
\end{equation}

\begin{figure*}
    \centering
    \includegraphics[width=0.49\textwidth]{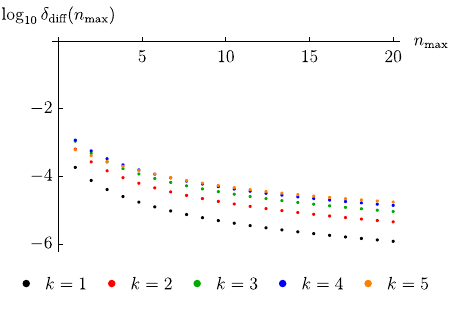}
    \hfill
    \includegraphics[width=0.49\textwidth]{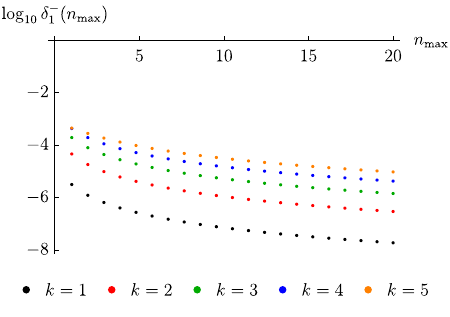}
    \caption{The plots show the relative error between the `accurate' numerically computed transverse channel modes (for different momenta $k$) and the ones computed from the $2n_\text{max}+2$ longitudinal spectrum modes. In the left panel, the relative error $\delta_\text{diff}(n_\text{max})=\abs{1-\widetilde\omega_\text{diff}/\omega_\text{diff}}$ is shown for the hydrodynamic diffusive mode $\widetilde\omega_\text{diff}$. In the right panel, the relative error $\delta_\text{1}^-(n_\text{max})=\abs{1-\widetilde\omega^-_{\text{gap},1}/\omega^-_{\text{gap},1}}$ is plotted for the first transverse gapped mode $\widetilde\omega^-_{\text{gap},1}$.}
     \label{fig:odd_error}
\end{figure*}

\begin{figure}
    \centering
    \includegraphics[width=\linewidth]{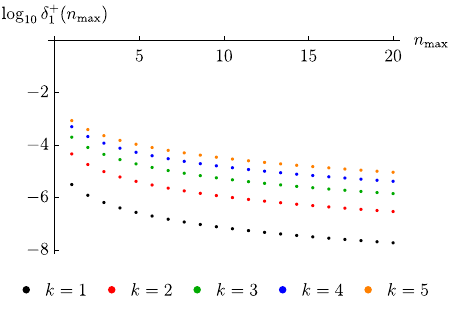}
    \caption{The plot shows the relative error between the `accurate' numerically computed first gapped mode in the longitudinal channel and the one computed from $2n_\text{max}+1$ transverse spectrum modes. As in Figure~\ref{fig:odd_error}, the relative error is defined as $\delta_1^+(n_\text{max})=\abs{1-\widetilde\omega^+_{\text{gap},1}/\omega^+_{\text{gap},1}}$. The constant $g_0^-$ was fixed using the hydrodynamic sound mode.}
    \label{fig:even_error}
\end{figure}

For our numerical demonstration, we have used the partial fraction expansion from Eqs.~\eqref{eq:partial_fractions} that includes a sum over $q=100$ gapped modes along with their mirror images. Figures \ref{fig:odd_error} and \ref{fig:even_error} show the errors, which rapidly decrease to below $0.1\%$.

\begin{figure}[ht!]
    \centering
    \includegraphics[width=1\linewidth]{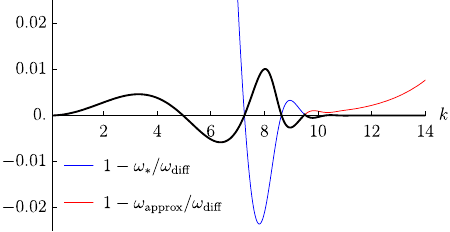}
    \caption{The figure depicts various relative errors that appear in the approximation of the hydrodynamic diffusive mode. In blue, we show the relative error that arises from using $\omega_*$. In red, we show the relative error that arises from using the approximation that includes the hydrodynamic sound mode, the asymptotic expansion \eqref{eq:WKB} and $\omega_*$. Their appropriate gluing is shown in black. The approximation fits the `exact' numerically computed diffusive mode within the accuracy of $1\%$.}
    \label{fig:approx_dif}
\end{figure}

Interestingly, a remarkably good approximation for the diffusive mode can be constructed using only the hydrodynamic sound mode dispersion relation, $\omega_*$ and the $k$-independent asymptotic expansion from Eq.~\eqref{eq:WKB}. We do this by setting $n_\text{max}=0$, i.e., by approximating all the gapped modes as in Eq.~\eqref{eq:w_approx}. In this way, we approximate the longitudinal channel spectral function by
\begin{equation}
    \rho^T(\omega)\approx\frac{H(\omega;0)}{H(0,0)}\frac{\mu \sinh\frac{\omega}{2T}}{\qty[1-\qty(\frac{\omega}{\omega_\text{sound}})^2]\qty[1-\qty(\frac{\omega}{\omega^*_\text{sound}})^2]}.
\end{equation}
The residues $r_n$ of $\rho^T(\omega)$ can be expressed analytically. However, for conciseness, we do not write them here. A partial fraction decomposition of $G^T_+(\omega)$ can then be performed in the sense of Eq.~\eqref{eq:Geven_partial}, using the two sound modes and the approximated gapped QNMs. The approximate diffusive mode $\omega_\text{approx}$ can then be computed numerically via Eq.~\eqref{eq:odd_compute}. Remarkably, it agrees with the exact diffusive mode to the accuracy of $1\%$ for $k\lesssim 9$. Beyond this point, $-\omega_*$ becomes a good approximation (cf.~Figure~\ref{fig:pole-skipping}). We can therefore glue the two expressions together to express (numerically) the hydrodynamic diffusive dispersion relation for all $k$ to the accuracy of $1\%$. This is shown in Figure~\ref{fig:approx_dif}. We have not attempted to analytically sum the partial fraction expansion in this case, nor did we tackle the problem in terms of a small-$k$ expansion. It would, however, be very interesting to see whether one can indeed express the diffusion constant analytically in terms of the sound mode, $\omega_*$ and the asymptotic large-$\omega$ behaviour of the QNMs alone.

\subsection{Beyond the Schwarzschild black brane}

In the previous subsection, we have demonstrated how dualities in the bulk constrain the thermal bulk QNM spectra and at the same time the boundary correlators. Here, we comment on more general setups, which can be treated in an analogous manner. Above, it was helpful for the analysis to be able to express the dualities in terms of `natural' electric and magnetic components of the electromagnetic field strength tensor and the Weyl tensor. For a CFT with a finite chemical potential dual to a charged Reissner-Nordstr\" om black brane, or in the linear axion model, such an analysis is significantly more involved, and we do not present it here. Nevertheless, we make several comments about both cases. Note that we did not verify the spectral duality relation for these cases explicitly. It has been recently verified, however, that the thermal product formula holds for correlators dual to a specific decoupled sector of perturbations \cite{Bhattacharya:2025vyi}, suggesting the validity of the analysis presented here.

Interestingly, the $\omega_* \rightarrow \infty$ limit in both the charged and the axion cases corresponds to the limit where the equilibrium expectation values of the corresponding operators vanish. The self-duality of the U$(1)$ current coincides with the vanishing of the charge density, whereas the self-duality of the metric perturbations coincides with the vanishing of the energy density, even in the presence of an event horizon. We suspect that this behaviour might be universal whenever such dualities are present.

\subsubsection{Finite chemical potential state and the Reissner-Nordstr\" om black brane}

We first focus on the finite charge density case. The bulk-boundary dictionary relating the bulk curvature tensors and boundary quantities (see Eqs.~\eqref{def:currents} and \eqref{def:sourced_holo}) can be checked to be valid in this case as well. The U$(1)$ current and the energy-momentum tensor now couple at the level of two-point functions. That is, the cross-correlator $\expval{TJ}\neq 0$, and the duality maps, expressed in Section~\ref{subsub:RN}, now act on some mixed correlators. While the explicit form of those correlators can be derived from the master functions (see Appendix~\ref{app:master}), we do not present them here. The corresponding 4$d$ QNM spectra were studied in a large number of works, including Refs.~\cite{Berti:2003ud,Davison:2011uk,Jansen:2017oag,Edalati:2010hk}.

For the sake of completeness, we write down the boundary thermodynamic relations (see e.g.~Ref.~\cite{Hartnoll:2009sz}). The boundary energy and charge density are\footnote{Note that, due to coupling between the metric and the Maxwell field, there is only one overall normalisation constant $\gamma$ supplemented by the Maxwell coupling $\kappa$.}
\begin{subequations}
\begin{align}
    \overline{\epsilon} &= \expval{T^{tt}}_0 = 2\gamma M,\\
    \overline{n} &= \expval{J^t}_0 =  \sqrt{2} \gamma \kappa  Q_e,
\end{align}    
\end{subequations}
and are related to the temperature $T$ and the chemical potential $\mu$ by the following relations:
\begin{subequations}
\begin{align}
    \frac{\overline{\epsilon}}{\gamma (4\pi T)^3}&=\frac{\kappa^2\rho_0}{2}\qty(\frac{\mu}{4\pi T})^2+\rho_0^3, \\ 
    \frac{\overline{n}}{\gamma (4\pi T)^2} &= -\kappa^2 \rho_0 \frac{\mu}{4\pi T},
\end{align}    
\end{subequations}
where $r_0=(4\pi T)\rho_0$ is the position of the outer event horizon, given by
\begin{equation}
    \rho_0 = \frac{1}{6}\qty(1+\sqrt{1+6 \qty(\frac{\kappa\mu}{4\pi T})^2}).
\end{equation}
In terms of the boundary variables, the corresponding algebraically special frequencies are given by 
\begin{equation}
    \omega_{(\pm)}= i \frac{\gamma k^4}{6 \bar \epsilon}\left(\frac{1}{2}\pm \sqrt{\frac{1}{4}+\frac{8}{9}\qty(\frac{\overline{n}}{\kappa\overline{\epsilon}})^2 k^2}\right)^{-1}.
\end{equation}

\subsubsection{The linear axion model}
\label{subsubsec:axion}
We now turn to the linear axion case. Like in the charged case, we do not provide explicit expressions for the boundary correlators on which the dualities of Section \ref{subsubsec:linear_axion0} act. For the sake of completeness, we write down the boundary thermodynamic variables in the axion case as well. The energy density and the temperature are related through
\begin{equation}
    \overline{\epsilon} = 2\gamma M = \gamma\frac{m_0 (m_0^2-m^2)}{2\sqrt{2}},
\end{equation}
with
\begin{equation}
    m_0 = \frac{4\pi T+\sqrt{6m^2+(4\pi T)^2}}{3\sqrt{2}},
\end{equation}
where we note that, in this case, additional counterterms are necessary in the expression for the boundary energy-momentum tensor \eqref{def:Tmunu}. The algebraically special frequency is 
\begin{equation}
    \omega_*=i \gamma \frac{k^4+k^2 m^2}{6\overline{\epsilon}},
\end{equation}
which agrees with the pole-skipping analysis of Ref.~\cite{Arean:2020eus}. The limit of vanishing energy density gives the self-dual limit of $\omega_*\rightarrow \infty$, where the duality equation \eqref{eq:DUALITY} reduces to the self-dual relation from Eq.~\eqref{eq:SELF_DUALITY}. This limit allows for explicit computations of all correlators and was discussed in detail in Ref.~\cite{Davison:2014lua}. At the self-dual point of the theory, we have
\begin{subequations}
\begin{align}
    m&=m_0,\\
    \overline{\epsilon} &= 0,\\
    T&=\frac{m_0}{2\sqrt{2}\pi},
\end{align}    
\end{subequations}
which enables us to directly compute $S(\omega)$:
\begin{equation}
    S(\omega)=1+i\lambda \sinh\frac{\omega}{2T}.
\end{equation}
This explicitly verifies the spectral duality relation \eqref{eq:SDR}.
For both the energy-momentum sector and the U$(1)$ sector, we can express
\begin{equation}
    \lambda = -\frac{1}{\cos\frac{\pi}{2}\sqrt{1-\qty(\frac{k}{\pi T})^2}}.
\end{equation}
At zero momentum, the QNMs in the spectrum again limit to the multiples of the negative imaginary Matsubara frequencies in the same way as in the spectrum of the current correlators in a neutral state. This was discussed in Section~\ref{subsubsec:JJ} and in Ref.~\cite{Grozdanov:2024wgo}.

One can also write down the corresponding $S(\omega)$ for the two-point functions of the scalar primaries dual to the bulk axions, finding
\begin{equation}
    \lambda = -\frac{1}{\cosh\frac{\pi}{2}\sqrt{7+\qty(\frac{ k}{\pi T})^2}}.
\end{equation}
Here, the spectral duality relation relates the spectra of two theories related by a double-trace deformation, or, in holographic terms, theories with the Dirichlet and Neumann boundary conditions respectively. This is discussed at length in Ref.~\cite{GV:2025}. Note that in this case, {\em no} gapless hydrodynamic mode exists, and therefore the spectra do not reduce to have $\omega$ given by the multiples of the Matsubara frequencies at zero momentum.

\subsection{Algebraically special solutions and the boundary interpretation}
It can be readily seen from the definition of $S(\omega)$ (cf.~Eq.~\eqref{def:sasian}), as well as the analysis of Section~\ref{subsec:neutral}, that the most crucial piece of information extracted from the bulk physics is the algebraically special frequency $\omega_*$. While the definitions of the correlators have to be extracted from bulk physics as well, most of the interesting physics of the boundary QFT can be studied without those details. Therefore, it is important to understand the meaning of the algebraically special frequency from the boundary point of view. Discussions similar to the present here have appeared in Refs.~\cite{ioannisTalk,Bakas:2014kfa,Skenderis:2017dnh,BernardideFreitas:2014eoi}.

The algebraically special condition on the boundary is the condition that gives $G_+(\omega_*)=G_-(-\omega_*)=0$. To formulate this without the detailed definitions of the correlators, we suppose that the boundary energy-momentum and Cotton-York tensors are given by the electric and magnetic parts of the Weyl tensor, respectively (see Eqs.~\eqref{def:Tmunu} and \eqref{def:bdry_cotton}). We can then take the limit of Eq.~\eqref{eq:EBalgSpec} to the boundary in the presence of a set of external sources $J$ \cite{ioannisTalk}, finding\footnote{Here, the spatial traceless part of a tensor is defined through Eq.~\eqref{def:transverse_tensor}, using the regularised boundary projector \eqref{eq:boundary_proj}.}
\begin{equation}
    \label{eq:boundary_AS}  \expval{T^{\expval{ab}}}_{J}=\epsilon^{ta}{}_d C^{\expval{db}}. 
\end{equation}
In the context of this paper, Eq.~\eqref{eq:boundary_AS} can be interpreted as a defining equation for $\omega_*$ in the sense that its solutions will exist at $\omega = \pm\omega_*$. In terms of the retarded two-point correlators, schematically, this means that
\begin{equation}
\label{eq:boundary_AS_expanded}
    \frac{\delta\expval{T^{\expval{ab}}}_{J}}{\delta g_{cd}}\delta g_{cd}+\sum_i\frac{\delta\expval{T^{\expval{ab}}}_J}{\delta j_i}\delta j_i= \epsilon^{ta}{}_d \frac{\delta C^{\expval{db}}}{\delta g_{cd}} \delta g_{cd} ,
\end{equation}
where $j_i \in J$ are the sources of all the operators that couple to $T^{ab}$ at the level of two-point functions in a given state. The right-hand side of Eq.~\eqref{eq:boundary_AS_expanded} is evaluated through the definition of the Cotton-York tensor \eqref{def:cotton}. The following statement can then be made: If there exists such a configuration of the sources $\delta g$ and $\delta j$ (for a given $\omega$ and $k$), so that Eq.~\eqref{eq:boundary_AS_expanded} is satisfied, then that configuration corresponds to a bulk algebraically special solution. At present, it is unclear what this means purely in terms of the quantum or thermal field theory language and why such solutions should be `special' in a QFT.

\section{Discussion and conclusion}
In this paper, we have studied duality relations between the linearised perturbations in 4$d$ geometries with black holes. In the case of the asymptotically-anti-de Sitter Schwarzschild black brane, we have demonstrated how such dualities parallel the well-known electric-magnetic dualities, endowed with the algebraically special structure. The interplay between the dualities and the algebraic structure of spacetimes was explained and related to the phenomenon of pole skipping \cite{Grozdanov:2017ajz,Blake:2017ris,Blake:2018leo}. 

For the moment, it is unclear exactly how general are the dualities discussed in this paper. It is easy to construct models that break the electric-magnetic duality of the Maxwell field explicitly (see e.g.~Ref.~\cite{Myers:2010pk}). Presumably, various modifications of the action can easily break the duality structure in pure gravity as well. While we have focused on `maximally symmetric' black hole geometries, the analysis mostly depends on the two preferred null directions. Indeed, such dualities exist also on the backgrounds of Kerr black holes as well \cite{Chandrasekhar:1985kt}. It is therefore likely that such or similar duality structures will persist for any Petrov type-D background in the Einstein-Maxwell theory with massless scalars. Moreover, it would be interesting to understand why the algebraically special structure in the models considered is so symmetric between the channels (the longitudinal and transverse channel having each other's negative algebraically special frequencies), and how that relates to the dualities. Another open question is whether any analogue of the dualities discussed here can be constructed in higher than four bulk spacetime dimensions.

In asymptotically AdS spaces, the action of the dualities is quite subtle, and has dramatic consequences for the linear response theory in the boundary conformal field theory at finite temperature. Specifically, it implies a relation between the retarded correlators of the two dual channels. The spectral duality relation \cite{Grozdanov:2024wgo}, which can be derived from the general properties of spectra and dual correlators, then imposes very stringent constraints on those spectra of both channels. We have used this fact to derive several results. Most importantly, we have demonstrated how to derive the full spectrum in one channel from the knowledge of the spectrum in the other channel. The crucial ingredient are again the algebraically special frequencies provided by the bulk physics. It would be interesting to gain an analytic handle on the numerical computations of Section~\ref{subsubsec:numerics}, for example, to acquire a good analytic approximation of the hydrodynamic diffusive mode's dispersion relation as a function of the sound mode's dispersion relation. Another avenue for future pursuits would be to try and use the spectral duality relation in combination with known CFT techniques, such as OPEs and the thermal bootstrap \cite{Caron-Huot:2009ypo,Iliesiu:2018fao,Petkou:2018ynm,Karlsson:2019qfi,Karlsson:2022osn,Marchetto:2023xap,Ceplak:2024bja,Buric:2025anb}. 

While we have provided an identity \eqref{eq:boundary_AS_expanded} on the retarded correlators which defines the algebraically special frequency in terms of the boundary physics, a concrete and physical interpretation of this expression is far from clear. Furthermore, while the self-dual cases are interpreted as corresponding to the particle-vortex self-duality or its gravitational analogue, the cases with a finite $\omega_*$ also lack a clear interpretation in the language of (quantum) thermal field theories. A concrete and pressing question is for example why such a structure would emerge in a thermal CFT$_3$ in the first place, and why is it controlled by the algebraically special frequency. Time and time again, it has been the gravity side of the holographic duality that provided us with a clear physical picture in the bulk, leaving its implications for the dual QFT mysterious. One such persistent example is the precisely the pole-skipping phenomenon, and, since the contents of this work and Ref.~\cite{Grozdanov:2024wgo} are closely related to pole skipping, perhaps unsurprisingly, the intuitive physical reasons behind the spectral duality relation and all its many consequences for now remain a mystery as well.     

\section{Acknowledgements}
We thank Tim Adamo, Latham Boyle, Guri Buza, Richard Davison, Giorgio Frangi, Pavel Kovtun, James Lucietti, Ioannis Papadimitriou, Alexander Soloviev and Alexander Zhiboedov for many elucidating and fruitful discussions. The work of S.G. was supported by the STFC Ernest Rutherford Fellowship ST/T00388X/1. The work is also supported by the research programme P1-0402 and the project J7-60121 of Slovenian Research Agency (ARIS). M.V. is funded by STFC Studentship ST/X508366/1.

\appendix
\section{Treatement via the Newman-Penrose formalism}
\label{app:NP}
The Newman-Penrose (NP) formalism \cite{Newman:1961qr} is a reformulation of the 4$d$ Lorentzian geometry in a way that is adapted to a pair of preferred null vectors. As such, it is well-suited for exploiting the algebraic structure of spacetimes in order to simplify the equations of motion. Much of the discussion of Section~\ref{sec:setup}, specifically, the electric and magnetic variables, is a reformulation of the objects that naturally appear in the NP formalism. In its presentation, we follow the conventions of Ref.~\cite{Stephani:2003tm}.

\subsection{The formalism}
The core ingredient of the NP formalism is a coordinate-free tetrad basis of complex null vectors. At every point, we have two real null vectors, $\ell^a$ and $k^a$, and two complex null vectors in a conjugate pair, namely, $m^a$ and $\bar{m}^a$, where we use the overline notation for complex conjugation. This is in order to remain consistent with the standard literature on the NP formalism. We subject the null vectors to the following constraints:
\begin{subequations}
\begin{align}
    \ell^a k_a&=-1, \label{eq:NP_normalisation}\\
    m^a \bar{m}_a&=1,
\end{align}
\end{subequations}
with all the other inner products being zero. The metric can then be expressed as
\begin{equation}
    g_{ab}=-2k_{(a}\ell_{b)}+2m_{(a}\bar{m}_{b)},
\end{equation}
and the projector from Eq.~\eqref{def:projectors} is expressed as
\begin{equation}
    \Delta_{ab}=2 m_{(a}\bar{m}_{b)}.
\end{equation}
We demand that 
\begin{equation}
    \epsilon_{abcd}\ell^a k^b m^c \bar{m}^d=i
\end{equation}
and make contact with Section~\ref{sec:setup} by identifying
\begin{subequations}
\begin{align}
    \ell^a&=\ell^a_-,\\
    k^a&=\ell^a_+,
\end{align}
\end{subequations}
which is consistent with the orthonormality condition \eqref{eq:NP_normalisation}. The background values of the null tetrad with respect to the metric \eqref{def:background_metric} are conveniently (but not uniquely) chosen as
\begin{subequations}
\label{eq:tetrad}
\begin{align}
    \ell^a\partial_a&=\frac{1}{\sqrt{2f(r)}}\qty(\partial_t-f(r)\partial_r),\\
    k^a\partial_a&=\frac{1}{\sqrt{2f(r)}}\qty(\partial_t+f(r)\partial_r),\\
    m^a\partial_a&=\frac{e^{i\theta}}{\sqrt{2}r}\qty(\sqrt{1-K \chi^2}\partial_\chi-i\frac{1}{\chi}\partial_\phi),\\
    \bar{m}^a\partial_a&=\frac{e^{-i\theta}}{\sqrt{2}r}\qty(\sqrt{1-K \chi^2}\partial_\chi+i\frac{1}{\chi}\partial_\phi),
\end{align}
\end{subequations}
where $\theta$ is an arbitrary real parameter that can be chosen for convenience.

The primary focus is the Weyl tensor, which can be completely described by the five complex Weyl scalars\footnote{It should be emphasised that the Weyl scalars are not scalars in the proper sense as they depend on the choice of the basis.}
\begin{subequations}
\begin{align}
    \Psi_0&=C_{abcd}k^am^bk^cm^d,\\
    \Psi_1&=C_{abcd}k^a\ell^bk^cm^d,\\
    \Psi_2&=C_{abcd}k^am^b\bar{m}^c\ell^d,\\
    \Psi_3&=C_{abcd}k^a\ell^b\bar{m}^c\ell^d,\\
    \Psi_4&=C_{abcd}\bar{m}^al^b\bar{m}^c\ell^d.
\end{align}
\end{subequations}
The only non-zero background Weyl scalar corresponding to the metric \eqref{def:background_metric} is $\Psi_2$, which usually denotes the `Coloumb-like' part of the Weyl tensor
\begin{equation}
    \Psi_2=\frac{r^2 f''(r)-2r f'(r)+2 f(r)-2K}{12r^2}.
\end{equation}
Suppose now that we have a normalised spacelike vector $n^a$ with $n^a n_a=1$, and suppose that we have chosen the definitions of $\ell^a$ and $k^a$ so that
\begin{subequations}
\begin{align}
    n^a&=\frac{1}{\sqrt{2}}\qty(k^a-\ell^a),\\
    u^a&=\frac{1}{\sqrt{2}}\qty(k^a+\ell^a),
\end{align}
\end{subequations}
in accordance with the notation of Section~\ref{sec:dualities}. We use $n^a$ to decompose the Weyl tensor into its electric $E_{ab}$ and magnetic $B_{ab}$ parts as in Eq.~\eqref{def:EBWeyl}. Taking the perturbed values of the basis vectors into account, we then 
have
\begin{align}
    E_{ab}+i B_{ab}&=\Psi_0 \bar{m}_a \bar{m}_b\\&+\sqrt{2}\Psi_1 (\bar{m}_a u_b+u_a \bar{m}_b)\nonumber\\&+\Psi_2(m_a \bar{m}_b+\bar{m}_a m_b+2u_a u_b) \nonumber\\&+\sqrt{2}\Psi_3 (m_a u_b+ u_a m_b)\nonumber\\&+\Psi_4 m_a m_b. \nonumber
\end{align}
Holographically, this allows us to express the boundary energy-momentum tensor and Cotton-York tensors with Weyl scalars through Eqs.~\eqref{def:Tmunu} and \eqref{def:bdry_cotton}. Furthermore, we have
\begin{equation}
    E_{\expval{ab}}+i B_{\expval{ab}}=\Psi_0 \bar{m}_a \bar{m}_b+\Psi_4 m_a m_b, \label{eq:NP_EBpsi}
\end{equation}
meaning that the quantities $E_\pm$ and $B_\pm$ defined in Eqs.~\eqref{def:EB} are directly related to the Weyl scalars $\Psi_0$ and $\Psi_4$. 

We now consider the perturbations in the sense of Section~\ref{subsec:perturbations}. Using the harmonic decomposition of $E_{\expval{ab}}+i B_{\expval{ab}}$ (see Eq.~\eqref{def:EB}), it is straightforward to show that
\begin{subequations}
\label{eq:NP_psi0psi4}
\begin{align}
    \Psi_0&=\frac{E_+-B_++i(E_-+B_-)}{r^2}\Upsilon(\chi,\phi)e^{-i\omega t}, \\ 
    \Psi_4&=\frac{E_++B_+-i(E_--B_-)}{r^2}\bar{\Upsilon}(\chi,\phi)e^{-i\omega t},
\end{align}
\end{subequations}
where
\begin{equation}
    \Upsilon(\chi,\phi)= r^2 Y_{ab}m^a m^b. 
\end{equation}
Here, it is to be understood that we are only considering a specific mode $\mu$. It is now transparent that the formalism of electric and magnetic variables utilised in this paper is just a different point of view on the NP formalism. It is well-known from the literature, e.g. from Ref.~\cite{Szekeres:1965ux}, that $\Psi_0$ and $\Psi_4$ denote the ingoing and outgoing waves, respectively. This is then directly transferable to the discussion of Section~\ref{subsec:gravitational_waves}.

\subsection{Algebraically special spacetimes}
In this section, we use the NP formalism to list several well-known statements about the algebraic structure of spacetimes and connect the statements made in Section~\ref{sec:algebraical} with the ones that appear in related literature.

We begin by introducing the Lorentz transformations on the null basis $(\ell,k,m,\bar{m})$. Two of them are null rotations that leave $k^a$ invariant,
\begin{subequations}
\label{eq:NP_nulll}
\begin{align}
    \ell^a &\rightarrow \ell^a+ B\bar{m}^a+\bar{B}m^a+B\bar{B}k^a,\\
    k^a &\rightarrow k^a,\\
    m^a &\rightarrow m^a,
\end{align}
\end{subequations}
where $B$ is a complex parameter. The corresponding action on the Weyl scalars is
\begin{subequations}
\begin{align}
    \Psi_0&\rightarrow \Psi_0, \\
    \Psi_1&\rightarrow \Psi_1+\bar B \Psi_2,\\
    \Psi_2&\rightarrow \Psi_2+2\bar B\Psi_1+\bar B^2\Psi_0,\\
    \Psi_3&\rightarrow \Psi_3+3\bar B \Psi_2+3\bar B^2\Psi_1+\bar B^3\Psi_0,\label{eq:Psi3rotation}\\
    \Psi_4&\rightarrow\Psi_4+4\bar B\Psi_3+6\bar B^2\Psi_2+4\bar B^3\Psi_1+\bar B^4\Psi_0. \label{eq:Psi4rotation}
\end{align}
\end{subequations}
Analogously, there are two null rotations that leave $\ell^a$ invariant,
\begin{subequations}
\label{eq:NP_nullk}
\begin{align}
    \ell^ a &\rightarrow \ell^a,\\
    k^a &\rightarrow k^a+E \bar{m}^a+\bar{E}m^a+E\bar{E}\ell^a,\\
    m^a &\rightarrow m^a,
\end{align}
\end{subequations}
where $E$ is, again, a complex parameter. The action of these rotations on the Weyl scalars is
\begin{subequations}
\begin{align}
    \Psi_0&\rightarrow\Psi_0+4E\Psi_1+6E^2\Psi_2+4E^3\Psi_3+E^4\Psi_4,\\
    \Psi_1&\rightarrow \Psi_1+3E \Psi_2+3E^2\Psi_3+E^3\Psi_4,\\
    \Psi_2&\rightarrow \Psi_2+2E\Psi_3+E^2\Psi_4,\\
    \Psi_3&\rightarrow \Psi_3+E \Psi_4,\\
    \Psi_4&\rightarrow \Psi_4 .
\end{align}
\end{subequations}
The remaining two Lorentz transformations correspond to the rescalings of $\ell^a$ and $k^a$, and to the complex rotation of $m^a$. Those will not be of interest to us here.

The null direction $k^a$ is a principal null direction (PND), in the sense of Eq.~\eqref{def:PND}, if and only if \cite{Penrose:1960eq,Jordan:1961}
\begin{equation}
    \Psi_0 = 0.
\end{equation}
Furthermore, the null direction $k^a$ is degenerate, in the sense of Eq.~\eqref{def:degeneratePND}, if one can perform a null rotation \eqref{eq:NP_nulll} so that 
\begin{equation}
    \Psi_0=\Psi_1=0. \label{eq:NP_dPND1}
\end{equation}
Similarly, the null direction $\ell^a$ is a PND, 
\begin{equation}
    \Psi_4=0, \label{eq:NP_PND2}
\end{equation}
and it is degenerate if one can perform a null rotation \eqref{eq:NP_nullk}, so that
\begin{equation}
    \Psi_4=\Psi_3=0.\label{eq:NP_dPND2}
\end{equation}
Combining the conditions \eqref{eq:NP_dPND1} and \eqref{eq:NP_dPND2} with the expression \eqref{eq:NP_EBpsi}, we get the condition \eqref{eq:EBalgSpec}. Furthermore, by utilising Eq.~\eqref{eq:NP_psi0psi4}, we get Eq.~\eqref{eq:perturbed_conditions}.

We can find the principal null directions by performing a null rotation \eqref{eq:NP_nulll} and demanding that $\ell^a$ is a PND via Eq.~\eqref{eq:NP_PND2}. This amounts to solving
\begin{equation}
\Psi_4+4\bar B\Psi_3+6\bar B^2\Psi_2+4\bar B^3\Psi_1+\bar B^4\Psi_0 = 0 \label{eq:PND_solve}
\end{equation}
for $\bar B$. If the order of Eq.~\eqref{eq:PND_solve} is less than $4$, then $k^a$ is a PND. Specifically, if Eq.~\eqref{eq:PND_solve} is of order $4-n$, then $k^a$ is an $n$-fold degenerate PND. A completely analogous and equivalent construction can be made by rotating $k^a$ into a PND by using the null rotations \eqref{eq:NP_nullk}.

Suppose now that are working with a type-D background, which means that we have
\begin{equation}
    \Psi_0=\Psi_1=\Psi_3=\Psi_4=0, \qquad
    \Psi_2 \neq 0,    
\end{equation}
as is the case for the geometries of Eq.~\eqref{def:background_metric}. Clearly, $\ell^a$ and $k^a$ are both doubly degenerate PNDs. Perturbing the metric $g_{ab} \rightarrow g_{ab}+ \delta g_{ab}$ gives linear corrections to the Weyl scalars,
\begin{subequations}
\begin{align}
    \Psi_0 &\rightarrow \delta \Psi_0, \\
    \Psi_1 &\rightarrow \delta \Psi_1, \\
    \Psi_2 &\rightarrow \Psi_2 + \delta \Psi_2,\\
    \Psi_3 &\rightarrow \delta \Psi_3,\\
    \Psi_4 &\rightarrow \delta \Psi_4,
\end{align}
\end{subequations}
where the $\delta$ symbol indicates a linearly small quantity. We note that, in the case at hand, $\Psi_0$ and $\Psi_4$ are independent of the perturbed null basis and can be reliably computed using only the background tetrad \eqref{eq:tetrad}.

We now attempt to diagnose the algebraic type of the perturbed spacetime. First, we consider the case of $\Psi_4 \neq 0$. Then we can try to find a PND by a null rotation \eqref{eq:NP_nulll}. In order for the corrections to the null vector $\ell^a$ to be linearly small, we have to take the rotation parameter $B$ to be linearly small as well. However, since $\Psi_3$ is linearly small, Eq.~\eqref{eq:Psi4rotation} leaves $\Psi_4$ invariant at the linear order. Analogous argument can be made for the vector $k^a$ and the Weyl scalar $\Psi_0$. The equations for finding PNDs therefore have no solutions on the perturbed spacetime. Now we consider the case of $\Psi_4=0$. We can use the null rotation \eqref{eq:NP_nulll} to set $\Psi_3$ to zero via Eq.~\eqref{eq:Psi4rotation}. This is possible because $\Psi_2$ is non-zero. In this case, $\ell^a$ is a doubly degenerate PND, and the perturbed spacetime can be considered algebraically special. This proves the statement made in Section~\ref{subsec:algebraically_linear}.

There exists extensive literature on the Goldberg-Sachs theorem within the framework of NP formalism, including the original paper \cite{Newman:1961qr}. Here, we only mention that the shear-free condition of a null geodesic \eqref{eq:shear-free} is translated into the following expression:
\begin{equation}
    m^a m^b\nabla_{a}\ell_{b}=0.
\end{equation}

\subsection{Master equations}
In the language of variables introduced in Section \ref{subsec:perturbations}, it is very easy to compactly write the equations of motion from Eqs.~\eqref{eq:evenEoM} and \eqref{eq:oddEoM} as
\begin{align}
    F(-i \partial_t,r)\qty[r^3 \psi_0(t,r,\chi,\phi)]&=0,\\
    F(i\partial_t,r)\qty[r^3 \psi_4(t,r,\chi,\phi)]&=0,
\end{align}
with the operator $F$ defined as in Eq.~\eqref{def:F}. Because the Weyl scalars are complex-valued, we have substituted $\omega \rightarrow i \partial_t$ to avoid any confusion. A similar form of the equations of motion to the one above was originally studied by Chandrasekhar in Ref.~\cite{Chandrasekhar:1985kt}.

\section{Master functions and maps between them}
\label{app:master}
In this appendix, we write down all the master functions that, by definition, obey the master equation \eqref{eq:master} in all the cases studied in this paper. Furthermore, we derive and prove the duality relations of Eqs.~\eqref{eq:dualityEMmetric}.
Throughout, we work in the Regge-Wheeler gauge with metric perturbations given by
\begin{subequations}
\label{eq:MSTR_metric}
\begin{align}
    \delta g_{ab} &= h_{ab}(r) Y e^{-i \omega t}, \\
    \delta g_{aB} &= v_a(r) X_B e^{-i \omega t},\\
    \delta g_{AB} &= r^2 w(r) \gamma_{AB} Y e^{-i \omega t}.
\end{align}
\end{subequations}
The electromagnetic field is perturbed by
\begin{subequations}
\label{eq:MSTR_maxwell}
\begin{align}
    \delta A_a&= A^+_a(r) Y e^{-i \omega t},\\
    \delta A_A&= A^-(r) X_A e^{-i \omega t}.
\end{align}
\end{subequations}
While the resulting variables are not manifestly gauge invariant since we expressed them in the Regge-Wheeler gauge, their gauge-invariant forms do exist in literature \cite{Martel:2005ir,Kodama:2003jz} and can be derived by gauge transformations of Eqs.~\eqref{eq:MSTR_metric} and \eqref{eq:MSTR_maxwell}.

\subsection{The Schwarzschild black hole}
We first consider the neutral Schwarzschild case of Section~\ref{subsubsec:schw}. The master functions for the Maxwell field, associated with the potential
\begin{equation}
    V^\text{Maxwell}_\pm = \mu \frac{f(r)}{r^2}
\end{equation}
are defined as
\begin{subequations}
\label{eq:MSTR_chiDef}
\begin{align}
\chi_+(r)&=r^2\qty(i\omega A_r^+(r)+\partial_r A^+_t(r)),\\
\chi_-(r)&=A^-(r).
\end{align}
\end{subequations}
The master equations for the metric form a Darboux pair, controlled by
\begin{align}
    \omega_* &= i \frac{\mu(\mu-2K)}{12M},\\
    \phi(r) &= \mu -2K +\frac{6M}{r},
\end{align}
and are given by
\begin{subequations}
\label{eq:MSTR_metridDef}
\begin{align}
    \psi_+(r)&=\frac{1}{\phi(r)}\qty(\frac{f(r)h_{tr}(r)}{i\omega}+r w(r)),\\
    \psi_-(r)&=r\qty(v_t'(r)+i \omega v_r(r))-2v_t(r).
\end{align}
\end{subequations}

\subsubsection{Proof of the duality relation}
Suppose we work on shell, i.e., with the functions that solve the equations of motion. Within a specific channel of perturbations, we can find maps between different variables using first-order differential operators. To illustrate this, suppose that $X(r)$ and $Z(r)$ are two such variables, and suppose that $X(r)$ obeys the equation of motion
\begin{equation}
    X''(r)+p_X(r)X'(r)+q_X X(r)=0. \label{eq:MSTR_Xeom}
\end{equation}
On shell, $Z(r)$ can be expressed as a linear combination of $X(r)$ and $X'(r)$,
\begin{equation}
    Z(r)=\alpha(r)X(r)+\beta(r)X'(r), \label{eq:MSTR_Zdef}
\end{equation}
since higher derivatives of $X(r)$ can be eliminated by using the equation of motion for $X(r)$. Taking the derivative of Eq.~\eqref{eq:MSTR_Zdef} and eliminating $X''(r)$, we can write
\begin{equation}
    \underline{Z}=\mathbf{M}_{ZX}\underline{X}, \label{eq:MSTR_Mdef}
\end{equation}
where $M_{ZX}$ is a $2\times 2$ matrix
\begin{equation}
    \mathbf{M}_{ZX}=\mqty(\alpha(r) & \beta(r)\\ \alpha'(r)-\beta(r)q_X(r) & \alpha(r)+\beta'(r)-\beta(r)p_X(r)),
\end{equation}
and the underlined quantities are column vectors defined as
\begin{align}
    \underline{X}&=\mqty(X(r) \\ X'(r)),\\
    \underline{Z}&=\mqty(Z(r) \\ Z'(r)).
\end{align}
The inverse map is induced by $\mathbf{M}_{XZ}=\mathbf{M}_{ZX}^{-1}$, and the equation of motion for $Z(r)$ is given by
\begin{equation}
    Z''(r)+p_Z(r)Z'(r)+q_Z Z(r)=0,
\end{equation}
where $p_Z(r)$ and $q_Z(r)$ are acquired by taking the derivative of Eq.~\eqref{eq:MSTR_Mdef} and eliminating $X(r)$ and all of its derivatives. The maps between variables and their respective $2$-nd order equations of motion are therefore encoded in $r$-dependent $2\times 2$ matrices.

In the case at hand, one set of relevant gauge invariant variables are the master functions $\psi_\pm(r)$ as defined in Eq.~\eqref{eq:MSTR_metridDef}. The other set are the variables $E_\pm(r)$ and $B_\pm(r)$ as introduced in Eq.~\eqref{def:EB}. We will denote the maps between them in the following way,
\begin{subequations}
\begin{align}
    \underline{E}_++\underline{B}_+&=\mathbf{M}_+(\omega) \underline{\psi}_+, \\
    \underline{E}_+-\underline{B}_+&=-\mathbf{M}_+(-\omega) \underline{\psi}_+, \\
    \underline{E}_--\underline{B}_-&=\mathbf{M}_-(\omega) \underline{\psi}_-, \\
    \underline{E}_-+\underline{B}_-&=\mathbf{M}_-(-\omega) \underline{\psi}_-.
\end{align}
\end{subequations}
Here, we took into account the fact that $\psi_\pm(r)$ satisfy time-reversal-invariant master equations \eqref{eq:master}, and that the variables of Eqs.~\eqref{eq:evenEoM} and \eqref{eq:oddEoM} satisfy pairwise time reversed equations of each other. The matrices $\mathbf{M}_\pm(\omega)$ are induced, in the sense described above, by the following maps:
\begin{subequations}
\begin{align}
    E_++B_+&=\qty[\Sigma^+(r)+\frac{r V_+(r)}{2i\omega f(r)}]\psi_+(r)\nonumber\\
    &+\frac{\Sigma^+(r)f(r)}{i\omega}\psi'_+(r), \\
    E_--B_-&=\qty[\Sigma^-(r)+\frac{r V_-(r)}{2i\omega f(r)}]\psi_-(r)\nonumber\\
    &+\frac{\Sigma^-(r)f(r)}{i\omega}\psi'_-(r),
\end{align}
\end{subequations}
where
\begin{subequations}
\begin{align}
-\Sigma^+(r)\phi(r)f(r)r^2 &=6M^2+r^2(2K-\mu)\Xi(r) \nonumber \\
&-Mr(6\Xi(r)-3\mu+2r^2\Lambda),\\
2rf(r)\Sigma^-(r)&=2r\Xi(r)-6M,
\end{align}
with
\begin{equation}
    \Xi(r)=K+i\omega r.
\end{equation}
\end{subequations}
The potentials $V_\pm(r)$ are defined by Eq.~\eqref{def:DarbouxPotentials}. The relations above are far from obvious and have to be derived from the perturbed Einstein's equations.

Given $(\underline{E}_++\underline{B}_+)$ as a solution of Eq.~\eqref{eq:Fexample}, we can now express
\begin{equation}
    \underline{E}_+-\underline{B}_+=-\mathbf{M}_+(-\omega)\mathbf{M}^{-1}_+(\omega)(\underline{E}_++\underline{B}_+), \label{eq:MSTReven}
\end{equation}
where the determinant of the map is, remarkably, independent of $r$:
\begin{equation}
    \det \left[ \mathbf{M}_+(-\omega)\mathbf{M}^{-1}_+(\omega)\right]=\frac{\omega_*+\omega}{\omega_*-\omega}.
\end{equation}
Similarly, in the odd channel, we have
\begin{equation}
  \underline{E}_--\underline{B}_-=\mathbf{M}_-(\omega)\mathbf{M}^{-1}_-(-\omega)(\underline{E}_-+\underline{B}_-), \label{eq:MSTRodd}
\end{equation}
where the determinant is again $r$-independent,
\begin{equation}
    \det \left[ \mathbf{M}_-(-\omega)\mathbf{M}^{-1}_-(\omega)\right] =\frac{\omega_*-\omega}{\omega_*+\omega}.
\end{equation}
This is a reflection of the fact that, at the algebraically special frequencies, one cannot map from variables that are zero when the full set of Einstein's equation is solved. We emphasise that, while the operators in Eqs.~\eqref{eq:MSTReven} and \eqref{eq:MSTRodd} denote non-trivial maps between the variables that satisfy the time-reversed equations of each other, they do not change the ingoing on the outgoing nature of the solution. This is because they provide a map between two different expression of the \emph{same} solution of the perturbed Einstein's equations.

Now we turn to the Darboux transformations of Section~\ref{subsec:Darboux}. The operators $L_\pm$ of Eq.~\eqref{def:Lpm} induce their respective $2 \times 2$ matrices, in the sense described above,
\begin{subequations}
\begin{align}
   \underline{\psi}_+=\mathbf{L}_+ \underline{\psi}_-,\\
   \underline{\psi}_-=\mathbf{L}_- \underline{\psi}_+.
\end{align}
\end{subequations}
As a consequence of working on shell, we have
\begin{equation}
    \mathbf{L}_+ \mathbf{L}_-=\mathbf{L}_- \mathbf{L}_+=(\omega^2-\omega_*^2) \mathbb{1}.
\end{equation}
Note that $\mathbf{L}_\pm$ have no $\omega$-dependence. The following can then be explicitly verified,
\begin{subequations}
\begin{align}
    \mathbf{M}_-(\omega)\mathbf{L}_-\mathbf{M}^{-1}_+(\omega) &=-i(\omega+\omega_*)\mathbb{1}, \\
    \mathbf{M}_+(\omega)\mathbf{L}_+\mathbf{M}^{-1}_-(\omega) &=i(\omega-\omega_*) \mathbb{1}.
\end{align}
\end{subequations}
These equations are precisely Eqs.~\eqref{eq:dualityEMmetric}, and the duality relation between the natural variables $E_\pm$ and $B_\pm$ is therefore proven.

\subsection{The Reissner-Nordstr\" om black hole}
For the Reissner-Nordstr\" om black hole considered in Section \ref{subsub:RN}, the perturbations form a Darboux pair in both the $(+)$ and the $(-)$ sectors, and are controlled by
\begin{align}
    \omega_{(\pm)}&=\frac{\omega_*}{\Delta_{(\pm)}},\\
    \phi^\qty(\pm)&=\mu-2K+\frac{6M}{r}\Delta^\qty(\pm),
\end{align}
where
\begin{subequations}
\begin{align}
Q&=\frac{Q_e}{3M},\\
\Delta^\qty(\pm)&=\frac{1}{2}\qty[1\pm\sqrt{1+4Q^2(\mu-2K)}],
\end{align}
\end{subequations}
The corresponding odd-channel master functions are
\begin{equation}
    \psi_-^\qty(\pm)=\rho_-(r)+\frac{\sqrt{2}\kappa \Delta^\qty(\pm)}{Q}  A^-(r),
\end{equation}
where
\begin{equation}
    \rho_-(r)=r\qty[v_t'(r)+i \omega v_r(r)]-2v_t(r)-\frac{2\sqrt{2}Q_e\kappa}{r} A^-(r).
\end{equation}
The corresponding even-channel master functions are
\begin{equation}
    \psi_+^\qty(\pm)=A_+(r)+\frac{\mu}{\sqrt{2}\kappa}\qty[\frac{Q_e}{r}-\frac{\Delta^\qty(\pm)}{2Q}
    ]\rho_+(r),
\end{equation}
where
\begin{subequations}
\begin{align}
    \rho_+(r)&=\frac{2(\mu-2K)}{\phi^\qty(+)(r)\phi^\qty(-)(r)}\qty(\frac{f(r)h_{tr}(r)}{i\omega}+r w(r)),\\
    A_+(r)&= r^2 \qty[i\omega A^+_r(r)+\partial_r A^+_t(r)]-\frac{\sqrt{2}Q_e}{\kappa}w(r).
\end{align}  
\end{subequations}
In the (appropriately regularised) zero-charge limit, functions $\psi^\qty(\pm)_\pm$ reduce to $\psi_\pm$ from Eq.~\eqref{eq:MSTR_metridDef} and $\chi_\pm$ from Eq.~\eqref{eq:MSTR_chiDef}.

\subsection{The linear axion model}
In the linear axion model considered in Section~\ref{subsubsec:axion}, there are two channels of perturbations that reduce to purely scalar perturbations in the $m\rightarrow 0$ limit. They are associated with the potential
\begin{equation}
    V_{y,z}=f(r)\frac{2k^2+m^2+6 r^2-2f(r)}{2r^2},
\end{equation}
with their respective master functions expressed as
\begin{subequations}
    \begin{align}
    \label{eq:MSTR_axionScalar}
    \varphi_y(r) = r \delta\Phi_y(r), \\
    \varphi_z(r) = r \delta\Phi_z(r), 
\end{align}
\end{subequations}
where we emphasise that $\varphi_y$ and $\varphi_z$ actually mix the scalar and metric perturbations, and the simple form of the definitions in Eq.~\eqref{eq:MSTR_axionScalar} is due to working in the Regge-Wheeler gauge. The channels that reduce to the purely metric tensor fluctuation channels in the $m\rightarrow 0$ limit form a Darboux pair controlled by
\begin{align}
    \omega_* &= i \frac{k^4+k^2m^2}{12M},\\
    \phi(r) &= k^2+m^2+\frac{6M}{r}.
\end{align}
The corresponding master functions are
\begin{subequations}
\begin{align}
    \psi_+(r)&=\frac{1}{\phi(r)}\qty(f(r) \frac{h_{tr}(r)}{i\omega}+r w(r))\nonumber\\&+\frac{mr}{ik(k^2+m^2)}\delta\Phi_z(r), \\
    \psi_-(r)&=r(v_t'(r)+i\omega v_r(r))-2v_t(r).
\end{align}
\end{subequations}

\newpage
\bibliography{references}

\end{document}